\documentclass[11pt, a4paper, oneside, reqno]{amsart}
\usepackage{comment}
\usepackage[usenames, dvipsnames]{color}
\definecolor{darkblue}{rgb}{0.0, 0.0, 0.45}
\definecolor{lightblue}{RGB}{240,248,255}
\definecolor{lightblue2}{rgb}{0.68, 0.85, 0.9}
\definecolor{lightcyan}{rgb}{0.88, 1.0, 1.0}
\definecolor{palepink}{rgb}{0.98, 0.85, 0.87}

\usepackage[colorlinks	= true,
raiselinks	= true,
linkcolor	= darkblue, %MidnightBlue,
citecolor	= Mahogany,
urlcolor	= ForestGreen,
pdfauthor	= {Peyman Mohajerin Esfahani},
pdftitle	= {},
pdfkeywords	= {},
pdfsubject	= {},
plainpages	= false]{hyperref}

\usepackage{dsfont,amssymb,amsmath,enumitem} %,etaremune,savesym}
\usepackage{amsfonts,dsfont,mathtools, mathrsfs,amsthm} 
\usepackage[amssymb, thickqspace]{SIunits}
\usepackage{fancyhdr,mdframed,nicefrac}

\usepackage[norelsize, ruled, vlined]{algorithm2e}
\usepackage{algcompatible}

\usepackage{epsfig}
\usepackage{graphicx}
\usepackage{float}
\usepackage{caption}
\usepackage{subcaption}

\usepackage{courier}

\usepackage{multirow}
\usepackage{bigstrut}

\allowdisplaybreaks
\date{\today}
\makeatletter
\def\@settitle{\begin{center}%
		\baselineskip14\p@\relax
		\normalfont\LARGE\scshape\bfseries
		\@title
	\end{center}%
}

\def\@setauthors{%
  \begingroup
  \def\thanks{\protect\thanks@warning}%
  \trivlist
  \centering\footnotesize \@topsep30\p@\relax
  \advance\@topsep by -\baselineskip
  \item\relax
  \author@andify\authors
  \def\\{\protect\linebreak}%
  \authors%
  \ifx\@empty\contribs
  \else
    ,\penalty-3 \space \@setcontribs
    \@closetoccontribs
  \fi
  \endtrivlist
  \endgroup
}

\makeatother
\makeatletter

\def\subsection{\@startsection{subsection}{2}%
	\z@{.5\linespacing\@plus.7\linespacing}{.5\linespacing}%
	{\normalfont\large\bfseries}}

\def\subsubsection{\@startsection{subsubsection}{3}%
	\z@{.5\linespacing\@plus.7\linespacing}{.5\linespacing}%
	{\normalfont\itshape}}

\usepackage{multirow}
\usepackage{bigstrut}
\usepackage{wrapfig}
\usepackage{caption}

\renewcommand{\le}{\leqslant}

\DeclareSymbolFont{symbolsC}{U}{pxsyc}{m}{n}

\usepackage{bm}
\usepackage{amsthm}

\usepackage{amssymb}
\usepackage{latexsym}
\usepackage{mathrsfs}  
\usepackage{bbm}
\usepackage{amsthm}

\usepackage{url}
\usepackage{color}
\usepackage{algorithm2e}
\usepackage{amsmath,bm}
\usepackage{subcaption}
\usepackage[makeroom]{cancel}

\usepackage{float}
\usepackage{lipsum}
\usepackage{dsfont}
\usepackage{rotating}
\usepackage{mathtools}
\usepackage{relsize}

\definecolor{myred}{rgb}{0, 0, 0}%pink
\definecolor{myred}{rgb}{0, 0.0, 0.0}%red
\definecolor{myred}{rgb}{0, 0, 0}%black

\title[A Continuous Stochastic Model for Non-Equilibrium Dense Gases]{
A Continuous Stochastic Model for Non-Equilibrium Dense Gases
}
 \author{Mohsen Sadr and M. Hossein Gorji}
 \thanks{Corresponding author: Mohsen Sadr}
 \thanks {Email: sadr@mathcces.rwth-aachen.de}
 \thanks{MATHCCES, Department of Mathematics, RWTH Aachen University, Schinkestrasse 2, D-52062 Aachen,
Germany.}

\date{December 20, 2017}
 
\begin{document}
\maketitle

\begin{abstract}
While accurate simulations of dense gas flows far from the equilibrium can be achieved by Direct Simulation  adapted to the Enskog equation, the significant computational demand required for collisions appears as a major constraint. 
In order to cope with that, an efficient yet accurate solution algorithm based on the Fokker-Planck approximation of the Enskog equation is devised in this paper; {the approximation is very much associated with} the Fokker-Planck model derived from the {Boltzmann equation}  by Jenny {\it et al}. [P. Jenny, M. Torrilhon, and S. Heinz, Journal of Computational Physics, 229, 1077 (2010)] and Gorji {\it et al}. [M. H. Gorji, M. Torrilhon, and P. Jenny, Journal of Fluid Mechanics, 680, 574 (2011)]. The idea behind these Fokker-Planck descriptions is to project the dynamics of discrete collisions implied by the molecular encounters into a set of continuous Markovian processes subject to the drift and  diffusion. Thereby, the evolution of particles representing the governing stochastic process becomes independent from each-other and thus very efficient numerical schemes can be constructed. By close inspection of the Enskog operator, it is observed that the dense gas effects contribute further to the advection of molecular quantities. That motivates a modelling approach where the dense gas corrections can be cast in the extra advection of particles. Therefore, the corresponding Fokker-Planck approximation is derived such that the evolution in the physical space accounts for the dense effects present in the pressure, stress tensor and heat fluxes. Hence the consistency between the devised Fokker-Planck approximation and the Enskog operator is shown for the velocity moments up to the heat fluxes. For validation studies, a homogeneous gas inside a box besides Fourier, Couette {and lid-driven cavity} flow setups are considered. 
The results based on the Fokker-Planck model are compared with respect to benchmark simulations, where good agreements are found for the flow field along with the transport properties.
\end{abstract}

\section{\label{sec:Intro}Introduction}
%%%%%%%%				WHY KINETIC THEORY
\noindent Conventional continuum models such as the Navier-Stokes-Fourier system (NSF) require closure assumptions for molecular fluxes. It is well known that these assumptions fail once a strong departure from the equilibrium is faced. In contrast, a higher level of closure can be attained through the notion of the velocity distribution function (VDF) whose dynamics is the subject of the kinetic models. In particular, the {\color{myred}Boltzmann equation} provides the evolution of the VDF for a dilute gas, resulting from the molecular chaos assumption.       However for high density flows, the dilute gas assumption underlying the {\color{myred}Boltzmann equation} breaks down and the physical dimensions of the molecules have to be taken into account. This scenario can be encountered in various gas flow phenomena such as unconventional gas reservoirs~\cite{Sander2017}, high pressure shock tubes~\cite{Petersen2001}, Sonoluminescence~\cite{brenner2002single} and interface of liquid-vapor at supercritical pressures~\cite{DAHMS20131667,Dahms20153648,DAHMS20151587}. In order to include dense effects in the VDF transport equation, Enskog has generalized the Boltzmann operator such that the resulting operator comprises the distance and correlation between the two colliding molecules \cite{enskog1922kinetische, Chapman1953}. \\ \ \\
%%%%%%%%				WHY DENSE IS IMPORTANT? FOR DENSE GAS, WE NEED ENSKOG. 
\noindent In general, three methodologies can be considered for gas flow simulations based on kinetic models: direct solution methods, moment methods and particle {\color{myred}Monte Carlo} schemes. The direct solution method addresses the VDF transport equation as the governing partial differential equation (PDE). Therefore the solution of the VDF is sought using different techniques such as the discrete velocity method (DVM)~\cite{broadwell_1964} or spectral methods \cite{Pareschi2000}, among others. Yet due to the high dimensionality of the phase space, the memory requirement for direct methods may become significant.
The next category belongs to the moment approach, where the PDEs governing the evolution of macroscopic moments are devised. Various moment approaches exist ranging from modified NSF \cite{karniadakis2005microflows} to Grad based approaches such as R13 \cite{Torrilhon2013}. While here fast simulations compared to other two methodes can be achieved, the closure problem besides boundary conditions for high order velocity moments are major challenges. The {\color{myred}Monte Carlo} methods employ the evolution of particle based realizations of the VDF consistent with the governing kinetic equation. In particular, the Direct Simulation  (DSMC) pioneered by Bird~\cite{Bird1963} is the mainstream solution method for simulations of rarefied gas flows.  While DSMC has been shown to be consistent with the solution of the {\color{myred}Boltzmann equation}~\cite{Wagner1992}, dense operations at low Knudsen numbers and large statistical errors at low Mach numbers have to be dealt with. \\ \ \\
%%%%%%%%				WHY Monte Carlo. SOLUTIONS TO ENSKOG
Various generalizations of DSMC for the Enskog equation have been pursued so far, mainly the  Enskog Simulation Monte Carlo method (ESMC)  derived by {\color{myred}Montanero and Santos \cite{Montanero1996,montanero1997simulation,montanero1997viscometric} } and modified DSMC by Frezzotti \cite{Frezzotti1997}. In comparison to DSMC, the Enskog based {\color{myred}Monte Carlo} schemes are computationally more demanding due to the explicit search of colliding pairs that can extend to several cells. A less complicated alternative is given by the Consistent Boltzmann Algorithm (CBA) proposed in {\color{myred}Ref.}~\cite{Alexander1995}, where its main difference with DSMC is that now the position update includes an extra streaming of the colliding pairs. The idea of the extra streaming comes from the fact that a certain minimum distance between the particles centers should be honored depending on the molecular potential under consideration. While CBA is based on a rather intuitive idea of the extra streaming, ESMC and Frezzotti's algorithms are directly deduced from the Enskog equation. Therefore in this study, converged results of ESMC are employed as the benchmark solutions of the Enskog equation. \\ \ \\
%%%%%%%%				WHY FP AT ALL?
\noindent A major drawback common among conventional DSMC based algorithms is that they require dense operations at low Knudsen numbers. This is due to the large number of collisions that have to be computed once the distribution becomes close to the equilibrium. While this deficiency roots in the jump processes governed by the Boltzmann/Enskog operators, {\color{myred}the Fokker-Planck (FP) approximation provides} a diffusion alternative in the form of It\={o} processes. Therefore, the resulting numerical schemes based on the FP equation do not account for collisions directly but via continuous stochastic differential equations (SDEs) and thus more efficient simulations can be expected; provided an accurate time integration scheme. The idea of the FP kinetic models dates back to the prominent works of Lebowitz {\it et al.}~\cite{Lebowitz1960} and Pawula~\cite{Pawula1967}. In the context of rarefied gas flow simulations, Jenny {\it et al.}~\cite{Jenny2010} have devised an efficient particle {\color{myred}Monte Carlo} scheme based on the FP approximation of the {\color{myred}Boltzmann equation}. Yet, since their FP model employed a simple Langevin closure for the drift and diffusion, only correct viscosity could be achieved. As a natural continuation of the mentioned study, Gorji {\it et al.}~\cite{Gorji2011} have devised a FP model with the cubic drift term such that relaxation of moments up to the heat fluxes consistent with the {\color{myred}Boltzmann equation}, could be honored. Since then, the cubic FP model has been validated for various scenarios ranging from the counter Fourier heat fluxes in the lid-driven cavity~\cite{Gorji2014,Gorji2015} to hypersonic flows~\cite{Gorji2015a,pfeiffer2017adaptive}. Moreover, the cubic FP model has been generalized to account for extra physical phenomena arising in polyatomic gases~\cite{Gorji2013} and mixtures~\cite{Gorji2012}. Note that several interesting alternatives to the cubic FP model have been developed by other research groups \cite{fei2013diffusive,Mathiaud2016,Bogomolov2016,Singh2016}, where the accuracy and efficiency of their models should be assessed in separate studies. \\ \ \\
%%%%%%%%				WHY DFP?
\noindent Motivated by the success of the FP based kinetic models for rarefied dilute gases, this work focuses on the extension of the cubic FP model for simulations of non-equilibrium gas flows including dense effects. To that extend, we propose a cubic FP model which results in accurate moment systems for the pressure, stress tensor and heat fluxes consistent with the Enskog equation. Naturally, the original cubic FP model proposed in {\color{myred}Ref.}~\cite{Gorji2011} can be recovered from the new model, passing to the low density limit.
In the following, i.e. Section~\ref{sec:theory} the Enskog equation together with its corresponding Taylor expansion around the Boltzmann operator are reviewed. Then in Section~\ref{sec:FP}, the dense FP model comprising an additional advection in the physical space is derived, in which accurate moments up to the heat fluxes are obtained. Section~\ref{sec:solution_algorithm} deals with a particle based solution algorithm appropriate for the dense FP model. In Section~\ref{sec:results}, numerical simulations of several test cases are carried out in order to assess the performance and accuracy of the devised model. Here comparisons with respect to CBA and ESMC are provided for evaluations. At the end in Section~\ref{sec:conclusion}, concluding remarks regarding the potentials of the proposed model along with a future outlook for possible improvements are briefly described.
\section{\label{sec:theory}Kinetic Theory}
\noindent In this section, the Enskog operator as the governing collision operator of a dense gas is reviewed. Due to the complexity of the operator, a simplified version resulting from its truncated expansion around the Boltzmann operator is considered. %Therefore, a dense FP model is devised where consistent moment systems for pressure, the stress tensor and the heat flux vector are met. 
%With that, a new FP model will be proposed such that dense effects could be obtained within the model. At the end of this section, the equivalent It$\hat{\text{o}}$ process is obtained using It$\hat{\text{o}}$'s lemma.
%Instead of solving PDE of Fokker-Planck equation, the equivalent SDEs are solved due to its simplicity when a particle method is deployed.
 
\subsection{\label{sec:enskog}Enskog Collision Operator}
\noindent Consider a gas composed of monatomic hard sphere molecules with the diameter $\sigma$ and mass $m$, whose velocities change in time due to binary collisions. The VDF $\mathcal{F}(\bm V,\bm x, t)$ quantifies the molecular distribution in the physical space $\bm x$, the velocity space $\bm V$ at each instant in time $t$, through
\begin{eqnarray}
\mathcal{F}(\bm V,\bm x,t)&=&\rho(\bm x,t)f(\bm V;\bm x,t),
\end{eqnarray}
where $f$ is the probability density of finding a molecule close to the velocity $\bm V$ and conditional on $\bm x-t$. Here $\rho$ is the gas density and is related to the number density $n$ by $\rho=nm$. \\ \ \\
\noindent Enskog has derived the operator which accounts for binary collisions in a dense gas through two modifications in the Boltzmann operator for hard sphere potentials. First, the position of colliding VDFs are distanced by $\sigma$ and second, the collision rate is increased by a factor $Y$ known as the pair correlation function. The latter is due to the reduction of available free space per unit of volume. Accordingly, the Enskog collision operator takes the form    
%\begin{widetext}
\begin{flalign}
S^{\text{Ensk}}
= \frac{1}{m}\int_{\mathbb{R}^3} \int_{0}^{2\pi} \int_{0}^{+\infty}  &
 \Big [ Y (\bm{x}+\frac{1}{2} \sigma \hat{\bm{k}}) \mathcal{F}(\bm{V}^*, \bm{x})\mathcal{F}(\bm{V}^*_1,\bm{x}+\sigma \hat{\bm{k}})
\nonumber \\
 &-Y (\bm{x}-\frac{1}{2} \sigma \hat{\bm{k}})\mathcal{F}(\bm{V}, \bm{x})\mathcal{F}(\bm{V}_1, \bm{x}-\sigma \hat{\bm{k}}) \Big ] 
g \hat{b} d\hat{b} d\hat{\epsilon}  d^3 \bm{V}_1.
\label{eq:enskog_op}
\end{flalign}
%\end{widetext}
Here $\hat{\bm{k}}$ is the normal vector connecting centers of two colliding hard spheres of velocities $\bm{V}$ and $\bm{V}_1$.
The superscript $^{*}$ indicates the post-collision state and the magnitude of the relative velocity is given by $g=|\bm{V}-\bm{V_1}|$. Furthermore, in order to determine the collision plane, $\hat{b}$ and $\hat{\epsilon} \in [0,2\pi]$ provide the impact parameter and the scattering angle, respectively. Also the abbreviation $d^3 \bm{V}=d {V}_1d {V}_2d {V}_3$ is used. For derivation and details of the {\color{myred}collision integral see  e.g. {\color{myred}Ref.}~\cite{Chapman1953}.} \\ \ \\
The Enskog operator requires a closure for the correlation function $Y$. Here the following approximation is considered
\begin{eqnarray}
Y \approx 1+ 0.625 n b + 0.2869  (nb)^2 + 0.1103  (nb)^3\ ,
\label{eq:Y}
\end{eqnarray}
where  $b=2 \pi \sigma^3/3$  is the hard sphere second Virial coefficient {\color{myred}~\cite{ree1967seventh}. While more rigorous formulation of $Y$ can be found based on the Carnahan-Starling equation \cite{carnahan1969equation}, for the considered range of $nb\le 1$ the above closure will be accurate enough. }%Note that for the range of $nb\le 0.5$ 

%%%%%%%%%%%%%%%%%%%%%%%%%%%%%%%%%%%%%%%%%%%%%%%%%
%%%%%%%%%%%%%%%%%%%%%%%%%%%%%%%%%%%%%%%%%%%%%%%%%
%%%%%%%%%%%%%%%%%%%%%%%%%%%%%%%%%%%%%%%%%%%%%%%%%

\subsection{Homogeneous Relaxation Rates\label{sec:homo_relax_rates}}
\noindent The consistency between the FP approximation and $S^{\text{Ensk}}$ should include the {\color{myred}kinetic} stress tensor 
\begin{eqnarray}
\pi_{ij}&=&\int_{\mathbb{R}^3}v^\prime_{\langle i}v^\prime_{j \rangle} \mathcal {F} d^3\bm V
\end{eqnarray}
and the {\color{myred}kinetic} heat fluxes
\begin{eqnarray}
q_{i}&=&\frac{1}{2}\int_{\mathbb{R}^3}v^\prime_{i}v^\prime_{j}v^\prime_{j} \mathcal {F} d^3\bm V,
\end{eqnarray}
along with the {\color{myred}kinetic} pressure $p$, the density $\rho$ and the bulk momentum $\rho \bm U$ with $\bm U$ being the bulk velocity. Here $v_i^\prime = V_{i} - U_{i}$ is the fluctuating velocity. Furthermore, the Einstein summation convention is employed and the subscript $\langle...\rangle$ denotes the deviatoric part of the tensor. \\ \ \\
\noindent In order to obtain closed form relaxation rates for different moments of $\mathcal{F}$, it is convenient to consider the decomposition 
\begin{eqnarray}
S^{\text{Ensk}} = Y S^{\text{Boltz}} + S_{\phi},
\label{eq:expansion_enskog}
\end{eqnarray}
where 
{\color{myred} $Y$ can be taken out of the collision integral for the homogeneous part
and} $S_{\phi} $ includes the spatial dependency of $\mathcal{F}$. 
%necessary to simplify the Enskog operator. By noting that $S^{\text{Enskog}}$ reduces to the Boltzmann operator $S^{\text{Boltz}}$ for space independent distributions, a Taylor expansion around $S^{\text{Boltz}}$ becomes attractive. To start, consider the decomposition
%By expanding Enskog operator in position using Taylor's theorem, one can derive\cite{Chapman1953}
%Therefore, no approximation is considered in Eq.~(\ref{eq:expansion_enskog}).
Therefore the corresponding rates follow
\begin{eqnarray}
\frac{\partial \pi_{ij}}{\partial t}{\color{myred}|_{coll}} &=&  Y\int_{\mathbb{R}^3} v^\prime_{\langle i} v^\prime_{ j \rangle} S^{\text{Boltz}} d^3 \bm{V} \label{relaxations_stress}\\
\text{and}  \ \ \ \ \ \frac{\partial q_{i}}{\partial t}{\color{myred}|_{coll}}  &= &\frac{Y}{2}\int_{\mathbb{R}^3} v^\prime_{ i} v^\prime_{ j}  v^\prime_{ j}  S^{\text{Boltz}} d^3 \bm{V}.
\label{eq:relaxations_heat}
\end{eqnarray}
By applying the monatomic Maxwell molecular model for $S^{\text{Boltz}}$ (see {\color{myred}Ref.}~\cite{C.Truesdell1980}), the corresponding rates for the Enskog operator become 
\begin{eqnarray}
\frac{\partial \pi_{ij}}{\partial t} {\color{myred}|_{coll}}&=&  - Y \frac{p}{\mu^{\text{kin}} } \pi_{ij}  \label{eq:relaxations_stress_enskog}\\
\textrm{and} \ \ \ \ \ \frac{\partial q_{i}}{\partial t} {\color{myred}|_{coll}} &= &- Y \frac{2}{3} \frac{p}{\mu^{\text{kin}} } q_i~,
\label{eq:relaxations_heat_enskog}
\end{eqnarray}
where $p=nkT$ is the ideal gas pressure with temperature $T$ and the Boltzmann constant $k$. The dynamic viscosity in the ideal gas limit is given by $\mu^{\text{kin}}$. \\ \ \\
{\color{myred}Here, the relaxation rates for $\bm \pi$ and $\bm q$ are calculated using the Maxwell model. More realistic decay rates should read the corresponding hard sphere expressions which can be found from {\color{myred}Ref.}~\cite{gupta2012automated}. However note that by linearizing the resulting hard sphere rates, the Maxwell ones can be obtained.} \\ \ \\
While the providing rates i.e. Eqs. (\ref{eq:relaxations_stress_enskog}) and (\ref{eq:relaxations_heat_enskog}) have to be reproduced by the FP model, further contributions due to the spatial dependency of $\mathcal{F}$ to the evolution of velocity moments have to be taken into account. In the following, these contributions are expressed in terms of integrals of $\mathcal{F}$.  Moreover, the total pressure tensor besides the total heat flux vector are defined.

%In design of dense FP model (DFP),  relaxation rates of Enskog equation Eq. (\ref{eq:relaxations_stress_enskog}) and Eq. (\ref{eq:relaxations_heat_enskog}) will be deployed to correct the velocity evolution. Please note that the ratio of relaxation rates in hydrodynamic limit provides Prandtl number. 

%%%%%%%%%%%%%%%%%%%%%%%%%%%%%%%%%%%%%%%%%%%%%%%%%%
%%%%%%%%%%%%%%%%%%%%%%%%%%%%%%%%%%%%%%%%%%%%%%%%%%
%%%%%%%%%%%%%%%%%%%%%%%%%%%%%%%%%%%%%%%%%%%%%%%%%%

\subsection{Collisional Transfer of Molecular Quantities}
\label{sec:Collisional_Transfer_of_Molecular_Quantities}
\noindent The main contribution of dense effects can be seen once the spatial dependency of the Enskog operator is taken into account. However, closed form equations for production terms resulting from the integrals of the form
\begin{eqnarray}
&&\int_{\mathbb{R}^3} \psi(\bm V) S^{\text{Ensk}} d^3 \bm{V} \label{eq:int}
\end{eqnarray}
can not be obtained unless an approximation of $S_{\phi}$ in Eq.~(\ref{eq:expansion_enskog}) is utilized. {\color{myred}By applying the Taylor expansion on $\mathcal{F}$ and $Y$ evaluated at $(\bm x\pm \sigma \bm k)$ and $(\bm x \pm \sigma \bm k/2)$, respectively,  the above integral can be simplified. Here a closure of the form}  %with respect to $\bm x$ %to  $\mathcal{F}=\mathcal{F}(\bm{V}, \bm{x})$ and $\mathcal{F}_1=\mathcal{F}(\bm{V}_1, \bm{x}\pm \sigma \bm k)$,
%In order to control the accuracy of the approximation in terms of $\sigma$, a Taylor expansion with respect to the position is typically employed. 
% Dense effects in collisions can be derived by taking velocity moments of Enskog equation and investigating total pressure tensor and heat fluxes. Instead of Eq.~(\ref{eq:enskog_op}), moments of expanded operator Eq.~(\ref{eq:expansion_enskog}) is utilized.
%However, instead of  Eq.~(\ref{eq:expansion_enskog}), moments of  Enskog collision operator which is derived based on middle point of colliding pair is preferred for simplicity. This implies transferring physical coordinate of Eq.~(\ref{eq:enskog_op}) to $\bm{x}-\bm{r}$ with $\bm{r}$ being the contact point of colliding particles.
%by keeping only first order derivatives with respect to $\bm x$, a closure of the form
\begin{eqnarray}
\int_{\mathbb{R}^3} \psi  S_{\phi} d^3 \bm{V} &= & - \dfrac{\partial \Psi_i}{\partial x_i},
\label{eq:moment_enskog_eq}
\end{eqnarray}
is sought. \\ \ \\
Following {\color{myred}Ref.}~\cite{Hirschfelder1963}, the resulting expression of $\bm \Psi$ by discarding all the spatial derivatives higher than first order, take the form
%\begin{widetext}
\begin{flalign}
\Psi_i &\approx 
\frac{Y \sigma}{2m} \int_{\mathbb{R}^3}  \int_{\mathbb{R}^3} \int_{0}^{2\pi} \int_{0}^{+\infty}  
 (\psi^*-\psi) 
  \mathcal{F} \mathcal{F}_1 
  k_i g \hat{b} d\hat{b} d\hat{\epsilon}  d^3 \bm{V}_1 d^3\bm{V}
  \nonumber \\
  &+
  \frac{Y \sigma^2}{4m} \int_{\mathbb{R}^3}  \int_{\mathbb{R}^3} \int_{0}^{2\pi} \int_{0}^{+\infty}  
 (\psi^*-\psi)
k_i k_j  \mathcal{F} \mathcal{F}_1  \frac{\partial}{\partial x_j} \ln (\frac{\mathcal{F}}{\mathcal{F}_1})
g \hat{b} d\hat{b} d\hat{\epsilon}  d^3 \bm{V}_1 d^3\bm{V}.
\label{eq:psi_phi}
\end{flalign}
%\end{widetext}
{\color{myred}Note that the next term in the Taylor expansion will have a $n\sigma ^3$ pre-factor which will result  in $nb\le 1$ constraint on the accuracy of the above-mentioned truncation.}
%can be sought as shown in Ref.~\cite{Hirschfelder1963}.
%
%
%it was shown in {\cite{Hirschfelder1963} that collisional contribution to change in distribution function becomes
%\begin{eqnarray}
%\int_{\mathbb{R}^3} \psi \big (  \frac{\partial \mathcal{F}}{\partial t} + v_i  \frac{\partial \mathcal{F}}{\partial x_i} \ \big ) d^3 \bm{V} 
%&= 
%- \dfrac{\partial \Psi^\phi_i}{\partial x_i},
%\label{eq:moment_enskog_eq}
%\end{eqnarray}
%where $\bm{\Psi^\phi}$ is 
%\begin{widetext}
%\begin{flalign}
%\Psi^\phi_i &= 
%\frac{Y \sigma}{2} \int_{\mathbb{R}^3}  \int_{\mathbb{R}^3} \int_{0}^{2\pi} \int_{0}^{+\infty}  
% (\psi^*-\psi)
%  \mathcal{F} \mathcal{F}_1 g \hat{b} d\hat{b} d\hat{\epsilon}  d^3 \bm{V}_1 d^3\bm{V}
%  \nonumber \\
%  &+
%  \frac{Y \sigma^2}{4} \int_{\mathbb{R}^3}  \int_{\mathbb{R}^3} \int_{0}^{2\pi} \int_{0}^{+\infty}  
% (\psi^*-\psi)
%k_i  \mathcal{F} \mathcal{F}_1  \frac{\partial}{\partial x_i} ln (\frac{\mathcal{F}}{\mathcal{F}_1})
%g \hat{b} d\hat{b} d\hat{\epsilon}  d^3 \bm{V}_1 d^3\bm{V}.
%\label{eq:psi_phi}
%\end{flalign}
%\end{widetext}
%
Even though simplifications are involved in Eqs.~\eqref{eq:moment_enskog_eq} and \eqref{eq:psi_phi}, still further closures are required for the gradient $\partial \ln (\mathcal{F}/\mathcal{F}_1)/\partial x_j$.
Following Chapmann and Cowling \cite{Chapman1953} the approximation $\partial  \ln (\mathcal{F}/\mathcal{F}_1)/\partial x_j \approx \partial \ln (\mathcal{F}^0/\mathcal{F}_1^0)/\partial x_j$ is adopted, where $\mathcal{F}^0$ reads the Maxwelllian VDF. \\ \ \\
Consider the Enskog equation for the evolution of $\mathcal{F}$
\begin{eqnarray}
\frac{\partial \mathcal{F}}{\partial t}+\frac{\partial}{\partial x_i} \left(V_i\mathcal {F}\right)&=&S^{\text{Ensk}}.
\end{eqnarray}
Therefore by applying the closures \eqref{eq:moment_enskog_eq}-\eqref{eq:psi_phi}, the resulting evolution PDEs for the bulk momentum $\rho\bm U$ and {\color{myred}internal energy} $e_s=c_vT$ become
\begin{eqnarray}
\label{eq:mom}
\frac{\partial }{\partial t}\left(\rho U_i\right)+\frac{\partial}{\partial x_j}\left(\rho U_iU_j+\pi^{tot}_{ij}+p^{tot}\delta_{ij}\right)&=&0
\end{eqnarray}
and 
\begin{eqnarray}
\label{eq:heat}
%\frac{\partial \rho e_s }{\partial t}+\frac{\partial}{\partial x_j}\left(\rho U_ie_s+q^{tot}_i\right)&=&-(\pi_{ij}^{tot}+p^{tot}\delta_{ij})\frac{\partial U_i}{\partial x_j}
\frac{\partial \rho e_s }{\partial t}+\frac{\partial}{\partial x_j}\left(\rho U_ie_s+q^{tot}_i\right)&=&-\pi_{ij}^{tot}\frac{\partial U_{\langle i}}{\partial x_{j\rangle}}
-p^{tot} \frac{\partial U_i}{\partial x_i}
\end{eqnarray}
with
%the velocity moments $\psi \in \{ mv^\prime_k, m v^\prime_k v^\prime_k/2 \}$  provides a relation between total transfer of conserved quantity and velocity moments of the VDF. Accordingly the total pressure, total stress tensor and total heat fluxes take the form
\begin{eqnarray}
\label{eq:pressure_enskog}
p^{\text{tot}}&=&nkT(1+nbY)-w \frac{\partial U_k}{\partial x_k},  \\
\pi^{\text{tot}}_{ij}  &=&  ( 1+2nbY/5) \pi_{ij}
- (6w/5)  \frac{\partial U_{\langle i}}{ \partial x_{j \rangle}}
~% \text{and}
\label{eq:pressure_tensor_enskog}
\end{eqnarray}
and
\begin{eqnarray}
q_i^{\text{tot}} &= &
(1+3nbY/5) q_i
- c_v w  \frac{\partial T}{\partial x_i}.
\label{eq:heat_conduction_enskog}
\end{eqnarray}
Here  $c_v = 3k/(2m)$ is the heat capacity of monatomic gas at constant volume and
\begin{eqnarray}
w &=& (nb)^2 Y \sqrt{mkT}/(\pi^{3/2} \sigma^2)
\label{eq:w}
\end{eqnarray}
is the {\color{myred}bulk viscosity}. \\ \ \\
The evolution equations \eqref{eq:mom}-\eqref{eq:heat} besides the homogeneous relaxations \eqref{eq:relaxations_stress_enskog}-\eqref{eq:relaxations_heat_enskog}, provide a basis for consistency relations between our dense gas FP model and the Enskog equation. In the next section, it is shown how such a FP model can be derived resulting in Eqs.~\eqref{eq:mom}-\eqref{eq:heat} together with Eqs.~\eqref{eq:relaxations_stress_enskog}-\eqref{eq:relaxations_heat_enskog}.
\section{ Dense Gas Fokker-Planck Model}\label{sec:FP}
\subsection{Review of the Cubic Fokker-Planck Model\label{sec:review_cubic_FP}}
\noindent Various Fokker-Planck approximations of the Boltzmann type Master equations have been pursued in several studies e.g. see Refs.~\cite{Pawula1967,Cercignani,Jenny2010,Gorji2011}. In Pawula's method, the Kramers-Moyal expansion is employed for the temporal variation of the distribution. Therefore, by truncating the expansion after the second order derivatives, the FP approximation of the governing Master equation is achieved \cite{Pawula1967}. Cercignani has considered the grazing limit of the Boltzmann operator which is relevant for the Coulomb type intermolecular interactions, where again the Boltzmann operator reduces to a FP type operator \cite{Cercignani}. \\ \ \\
\noindent The approach developed by Gorji and colleagues in Refs.~\cite{Jenny2010,Gorji2011} follows a different path. The main step here is to introduce an expansion of the drift with respect to the fluctuating velocity. Therefore by matching the velocity moments resulting from the FP operator with those arising from the Boltzmann operator, the coefficients of the expansion are found. In this approach, one can directly control the number of moments of the FP operator that match with those of the governing Master equation. A review of this approach for monatomic Maxwell molecules is provided in the following.\\ \ \\
The main idea behind the FP approximation is to express the Boltzmann operator by
\begin{eqnarray}
S^{\text{Boltz}}\approx S^{\text{FP}}&=&-\frac{\partial}{\partial V_i}(A_i\mathcal{F})+\frac{\partial ^2}{\partial V_i\partial V_j}(D_{ij}\mathcal{F}) \label{eq:FP}
\end{eqnarray}
with the drift $\bm A$ and the positive definite diffusion $\bm D$. Note that in general $\bm A$ and $\bm D$ may depend on $\bm V$ and moments of $\mathcal{F}$, resulting in a highly nonlinear operator $S^{\text{FP}}$. The closures for $\bm A$ and $\bm D$ then can be found from the consistency relation
\begin{eqnarray}
\label{eq:consist_boltz}
\int_{\mathbb{R}^3}S^{\text{Boltz}}\psi d^3\bm V&=&\int_{\mathbb{R}^3}S^{\text{FP}}\psi d^3\bm V
\end{eqnarray}
for $\psi \in \{1,V_i,V_iV_j,...,V_{i_1}V_{i_2}...V_{i_n}\}$. \\ \ \\
The simplest relevant closure is given by the Langevin model
\begin{eqnarray}
A_i&=&-\frac{1}{\tau}v^\prime_i \ \ \ \ \textrm{and} \ \ \ \ \ D_{ij}=\frac{kT}{\tau m}\delta_{ij} \label{eq:diffusion-FP},
\end{eqnarray}
where the time scale $\tau=2\mu^{\text{kin}}/p$ results in consistencies with the {\color{myred}Boltzmann equation} for $\psi \in \{1,V_i,V_iV_j\}$ (see e.g. {\color{myred}Ref.}~\cite{Jenny2010}). In order to extend the domain of consistent moments up to the heat fluxes, a drift expansion of the form
\begin{eqnarray} 
A_i&=&c_{ij}v^\prime_j+\gamma_i\left(v^\prime_jv^\prime_j-\frac{3kT}{m}\right)+\Lambda \left(v^\prime_iv^\prime_j v^\prime_j-\frac{2q_i}{\rho}\right)
\label{eq:cubic_ansatz_ideal_gas}
\end{eqnarray}
was introduced as the cubic FP model in {\color{myred}Ref.}~\cite{Gorji2011}. The consistency relation \eqref{eq:consist_boltz} subject to $\psi \in \{1,V_i,V_iV_j,V_iV_jV_j\}$ provides nine linear equations for the symmetric tensor $\bm c$ and the vector $\bm \gamma$. Note that $\Lambda$ is a model constant introduced for the stability of the drift, and the diffusion remains the same as Eq.~\eqref{eq:diffusion-FP} (see {\color{myred}Ref.}~\cite{Gorji2011} for details). \\ \ \\
While expansions to higher order moments is possible, already the cubic model has shown to be accurate enough for a wide range of rarefied gas flow simulations at not too large Knudsen numbers \cite{Gorji2014}. Even though larger departures from the equilibrium may lead to less accurate results from the FP approximation, these high Knudsen regimes can be easily treated by DSMC. Therefore the idea of combining FP and DSMC solution methods have been developed and validated \cite{Gorji2015a}. In the following, a generalization of the cubic FP model for the case of the Enskog operator is presented.

%% Approximation to the Boltzmann
%% Cubic Model
\subsection{Fokker-Planck Model for Dense Gases}
\noindent As it was already discussed in Section~\ref{sec:homo_relax_rates}, the Enskog operator resembles the {\color{myred}Boltzmann one} in the spatially independent setting. Since already a cubic FP model has been derived for the Boltzmann operator, the homogeneous limit therefore can be easily included in the FP framework. Yet by simple inspection of the FP operator i.e. Eq.~\eqref{eq:FP}, it can be seen that the production terms of the form \eqref{eq:moment_enskog_eq} can not be reproduced. One approach to derive the appropriate generalization of the FP model is then based on the extra streaming in the physical space. Therefore a FP equation of the form
\begin{eqnarray}
\frac{\partial \mathcal{F}}{\partial t} 
+ \frac{ \partial }{\partial x_i}(V_i \mathcal{F}  ) 
&=& - \frac{ \partial}{\partial V_i} (A_i \mathcal{F}  )
+\frac{\partial^2}{\partial V_i \partial V_j} 
       ( D_{ij} \mathcal{F}) - \frac{ \partial}{\partial x_i} (\hat{A}_i \mathcal{F}  )~
\label{eq:FP_dense}
\end{eqnarray}
is considered, where the extra drift in the physical space $\bm{\hat{A}}$ controls the streaming due to dense effects. It is precisely this term which results in the production terms of the form \eqref{eq:moment_enskog_eq}. \\ \ \\
Physically, compared to point particles in an ideal gas, the spatial extension of gas molecules at high densities contribute further to the transport of molecular quantities during collisions. Therefore, a dense gas model should admit an extra advection in the physical space such that correct molecular fluxes can be obtained. This extra streaming can be regarded as a correction to the kinetic velocity that the molecules would attain if their diameters vanish. Note that similar intuition has been applied in derivation of the CBA algorithm. \\ \ \\
%% Give an insight on kinetic velocity and true velocity
In the following, we study the relationship between the introduced streaming and evolution of different macroscopic moments. Hence a consistent framework for the closure of $\bm {\hat{A}}$ is sought. \\ \ \\
In order to provide closures for the velocity drift $\bm A$, the diffusion $\bm D$ and the streaming $\bm {\hat{A}}$, the consistency relations arising from Eqs.~\eqref{eq:mom}-\eqref{eq:heat} along with Eqs.~\eqref{eq:relaxations_stress_enskog}-\eqref{eq:relaxations_heat_enskog} will be derived. First, a linear drift model which can be seen as a generalization of the Langevin equation for the dense media is provided. Next, a cubic model controlling the evolution of moments up to the heat fluxes is proposed.

%%%%%%%%%%%%%%%%%%%%%%%%%%%%%%%%%%%%%%%%%%%%%%%%%%
%%%%%%%%%%%%%%%%%%%%%%%%%%%%%%%%%%%%%%%%%%%%%%%%%%
%%%%%%%%%%%%%%%%%%%%%%%%%%%%%%%%%%%%%%%%%%%%%%%%%%
\subsubsection{\label{sec:linear_DFP}Linear drift}
\noindent {\color{myred}Before proceeding} to the more complicated case of the cubic dense model, {\color{myred}let us} consider a linear drift closure for the FP model. To start, notice that the spatial homogeneous condition provides us with consistency relations for $\bm A$ and $\bm D$. Therefore using Eqs.~\eqref{eq:relaxations_stress_enskog}-\eqref{eq:relaxations_heat_enskog}, the following closure is obtained
\begin{eqnarray}
A_i&=&-\frac{1}{\tau}v^\prime_i \ \ \ \ \textrm{and} \ \ \ \ \ D_{ij}=\frac{kT}{\tau m}\delta_{ij} \label{eq:diffusion-dFP},
\end{eqnarray}
where the time scale now becomes $\tau=2\mu^{\text{kin}}/(pY)$. Note that the time scale becomes {\color{myred}$Y$} times shorter compared to the ideal gas case. It is easy to check that the provided Eq.~\eqref{eq:diffusion-dFP} fulfills the constraint \eqref{eq:relaxations_stress_enskog}. However, further consistencies arising from Eqs.~\eqref{eq:mom}-\eqref{eq:heat} require a closure for $\bm{\hat{A}}$.  First note that the conservation of mass requires that 
\begin{eqnarray}
\int_{\mathbb{R}^3}\hat{A}_i\mathcal{F}&=&0. \label{eq:cons-const}
\end{eqnarray}
The above constraint suggests a polynomial expansion of $\hat{A}$ with respect to the fluctuating velocity. Therefore the simplest closure can be proposed as
\begin{eqnarray}
\hat{A}_i&=& \alpha v^\prime_i, \label{eq:linear-closure-st}
\end{eqnarray}
 where $\alpha$ is the introduced constant. First note that Eq.~\eqref{eq:linear-closure-st} leads to \eqref{eq:cons-const} and thus the mass conservation is guaranteed. Furthermore, in the following it is explained that $\alpha$ in fact controls the deviation of the gas pressure from the ideal gas law. \\ \ \\ 
\noindent The first velocity moment of Eq.~(\ref{eq:FP_dense}) provides the equation for conservation of momentum. Multiplying both sides by $V_i$ and taking integral over the velocity space, it follows
%%
%\begin{eqnarray}
%\frac{\partial (\rho U_k)}{\partial t}  + \frac{\partial (\rho \langle M_i M_k \rangle)}{\partial x_i} = 
%- \frac{\partial }{\partial x_i} ( \rho \langle \hat{A}_i M_k \rangle )~.
%\end{eqnarray}
%%
%By decomposing the velocity to its mean and fluctuating part $\bm{v} = \bm{v}^\prime + \bm{U}$, it follows
%
\begin{eqnarray}
\frac{\partial }{\partial t}(\rho U_i)  +  \frac{\partial }{\partial x_j} (\rho  U_i U_j )
+
 \frac{\partial  }{\partial x_j} (\pi_{ij}+p\delta_{ij})
+ \frac{\partial }{\partial x_j} \int_{\mathbb{R}^3}\hat{A}_j V_i \mathcal{F} d^3\bm V=0~.
\label{eq:1_momentum}
\end{eqnarray}
Now by using the linear drift \eqref{eq:linear-closure-st}, the integral is decomposed into
\begin{eqnarray}
\int_{\mathbb{R}^3}\hat{A}_j V_i \mathcal{F} d^3\bm V&=&\alpha\ nkT\delta_{ij}+\alpha\pi_{ij}.
\end{eqnarray}
Thus by comparing the total pressure resulting from the Enskog equation i.e. Eq.~\eqref{eq:pressure_enskog}, $\alpha$ can be found
\begin{eqnarray}
\alpha&=&nbY-\frac{w}{nkT}\frac{\partial U_i}{\partial x_i}.
\end{eqnarray}
Note that further consistency relations for the total stress and the total heat fluxes can not be deduced since only a scalar coefficient is introduced in the ansatz \eqref{eq:linear-closure-st}. A higher order expansion of the extra streaming with respect to the fluctuating velocity is carried out in the following section, such that the mentioned consistencies are fulfilled at the same time.
%where $p_{ik}^{\text{kin}}=\rho \langle M_i^\prime M_k^\prime \rangle$ is the kinetic pressure tensor.
%Now, we only need to find $\bm{\alpha}$ such that total pressure of Eq.~(\ref{eq:1_momentum}) matches total pressure obtained from Enskog equation, i.e. Eq.~(\ref{eq:pressure_tensor_enskog}). This leads to a $9\times9$ linear system of equations,
%%
%\begin{eqnarray}
% \rho \alpha_{ij} \langle   M^\prime_j M^\prime_k \rangle 
%% + \gamma_i  \langle  M^\prime_j M^\prime_j M^\prime_k \rangle
%=
%p_{ik}^{\text{tot}}-p_{ik}^{\text{kin}}~.
%\label{eq:1_shear_stress_correction}
%\end{eqnarray}
% %  
\subsubsection{\label{sec:cubic_FP}Cubic drift}
\noindent Similar to the linear drift case described before, here we distinguish {\color{myred}both the homogeneous and inhomogeneous cases}. {\color{myred}Let us} consider the consistency relations for the relaxation of non-equilibrium moments up to the heat fluxes. Since Eqs.~\eqref{eq:relaxations_stress_enskog}-\eqref{eq:relaxations_heat_enskog} are very similar to the case of dilute gas (only a factor {\color{myred}$Y$} is multiplied), a same structure as the cubic FP model described in Section~\ref{sec:review_cubic_FP} can be adopted for $\bm A$ and $\bm D$. Yet for completeness here we provide the resulting constitutive equations for the coefficients involved in $\bm A$ and $\bm D$. \\ \ \\
\noindent First note that the diffusion remains the same as Eq.\eqref{eq:diffusion-dFP} with the modified time scale explained before. Since only moments up to the heat fluxes are concerned, the cubic ansatz \eqref{eq:cubic_ansatz_ideal_gas} is adopted. Yet the coefficients $\bm c$, $\bm \gamma$ and $\Lambda$ will follow slightly different closures in comparison to the dilute case.
For convenience, the abbreviation 
\begin{eqnarray}  
u_{i_1...i_n}^{(k)}&=&\frac{1}{\rho}\int_{\mathbb{R}^3}|\bm {v^\prime}|^kv^\prime_{i_1}v^\prime_{i_2}...v^\prime_{i_n}\mathcal{F}d^3\bm V
\end{eqnarray}
is adopted in the following derivations.
By taking different velocity moments the FP equation \eqref{eq:FP_dense} and fulfilling the relaxation equations \eqref{eq:relaxations_stress_enskog}-\eqref{eq:relaxations_heat_enskog}, the constitutive equations 
\begin{eqnarray}
c_{ik}u^{(0)}_{kj}+c_{jk}u^{(0)}_{ki}+\gamma_iu^{(2)}_{j}+\gamma_ju^{(2)}_{i}&=&-2\Lambda u^{(2)}_{ij}  \label{eq:sys_cubic_FP_1}\ \ \ \ \ \textrm{and} \\ 
c_{ij}u^{(2)}_j+2c_{jk}u^{(0)}_{ijk}+\gamma_i(u^{(4)}-(u^{(2)})^2) && \nonumber \\
+2\gamma_j(u^{(2)}_{ij}-u^{(2)}u_{ij})&=&-\Lambda (3u^{(4)}_i-u^{(2)}_iu^{(2)}-2u^{(2)}_ju^{(0)}_{ij}) \nonumber \\
&&+\frac{5}{6}\frac{Yp}{\mu^{\text{kin}}}q_i \label{eq:sys_cubic_FP_2},
\end{eqnarray}
can be obtained. Similar to the dilute gas cubic model, the stability of the drift is controlled by the cubic term \cite{Gorji2011}
\begin{eqnarray}
\label{eq:lambda}
\Lambda=-\frac{|\det (u_{ij})|}{\tau (u^{(2)})^4}.
\end{eqnarray}
Equations \eqref{eq:sys_cubic_FP_1}-\eqref{eq:sys_cubic_FP_2} provide coefficients for the cubic drift term $\bm A$, where besides the expression \eqref{eq:diffusion-dFP} for $\bm D$, close the drift and diffusion in the velocity space. \\ \ \\
More challenging is to devise the extra streaming which leads to the moment system similar to those arising from the Enskog equation i.e. Eqs.~\eqref{eq:mom}-\eqref{eq:heat}. In analogy to the cubic drift of the velocity i.e. $\bm A$, the cubic ansatz
\begin{eqnarray}
\hat{A}_i &=& \hat{c}_{ij} v^\prime_j + \hat{\gamma}_i \left(v^\prime_j v^\prime_j - \frac{3kT}{m}\right) 
+ \hat{\Lambda} \Big ( v^\prime_i v^\prime_j v^\prime_j -  \frac{2q_i}{\rho} \Big )
\label{eq:2nd_A}
\end{eqnarray}
 is adopted for $\hat{\bm A}$. First note that the cubic coefficient $\hat{\Lambda}$ is chosen as a negative small number
\begin{eqnarray}
 \hat{\Lambda} = - \epsilon \frac{nbY}{k T/m }~,
\label{eq:Lambda*}
\end{eqnarray}
with $\epsilon=10^{-3}$, for stability of the extra streaming \cite{Risken1989,Gorji2011}.
Furthermore, in order to derive constitutive equations for the coefficients $\hat{\bm c}$, $\hat{\bm \gamma}$, the evolution equations for mass, momentum and energy are considered: 
\begin{enumerate}
\item  {\it Mass Conservation}: Since the introduced expansion \eqref{eq:2nd_A} has zero expectation, the mass conservation remains intact. More precisely due to
\begin{eqnarray}
\int_{\mathbb{R}^3}\hat{A}_i\mathcal{F}d^3\bm V&=&0,
\end{eqnarray}
the conservation of mass
\begin{eqnarray}
\frac{\partial \rho}{\partial t}+\frac{\partial}{\partial x_i}(\rho U_i)&=&0
\end{eqnarray}
can be derived by taking the integral of the proposed dense FP equation \eqref{eq:FP_dense} over the velocity space.
\item  {\it Momentum Conservation}: By taking the first order moment of Eq.~\eqref{eq:FP_dense} and replacing $\hat{\bm A}$ ansatz, we find
\begin{eqnarray}
\frac{\partial }{\partial t}(\rho U_i)  +  \frac{\partial }{\partial x_j} (\rho  U_i U_j )
+
 \frac{\partial  }{\partial x_j} (\pi_{ij}+p\delta_{ij})&& \nonumber \\
+ \frac{\partial }{\partial x_j} \left(\hat{c}_{jk}\pi_{ik}+\hat{c}_{ji}p+2\hat{\gamma}_jq_i+\rho \hat{\Lambda} u^{(2)}_{ij}\right)&=&0~.
%\label{eq:1_momentum}
\end{eqnarray}
In order to fulfill the momentum conservation resulting from the Enskog equation i.e. \eqref{eq:mom}, nine equations
\begin{eqnarray}
\label{eq:const_stress_total}
\hat{c}_{jk}\pi_{ik}+\hat{c}_{ji}p+2\hat{\gamma}_jq_i&=&-\rho \hat{\Lambda} u^{(2)}_{ij} \nonumber \\
&&+nbY(p\delta_{ij}+2/5\pi_{ij})-w\left(\frac{\partial U_k}{\partial x_k}\delta_{ij}+\frac{5}{6}\frac{\partial U_{\langle i}}{ \partial x_{j \rangle}}\right)
%\left(nbYp-{w}\frac{\partial U_k}{\partial x_k}\right)\delta_{ij} %\nonumber \\
%+ \frac{2nbY}{5} \pi_{ij}
%- \frac{5w}{6}  \frac{\partial U_{\langle i}}{ \partial x_{j \rangle}}
\end{eqnarray}
can be obtained for $\hat{\bm c}$ and $\hat{\bm \gamma}$. Yet to find the three remaining equations to close the system for $\hat{\bm c}$ and $\hat{\bm \gamma}$, the energy conservation has to be taken into account.
\item {\it Energy Conservation}: The evolution of {\color{myred}internal energy} according to the dense FP equation \eqref{eq:FP_dense} follows
\begin{eqnarray}
\frac{\partial \rho e_s }{\partial t}+\frac{\partial}{\partial x_j}\left(\rho U_ie_s+q_i\right)+\frac{1}{2}\frac{\partial}{\partial x_j}\int_{\mathbb{R}^3}\hat{A}_jv^\prime_kv^\prime_k\mathcal{F}d^3\bm V&=&-\pi^{tot}_{ij}\frac{\partial U_{\langle i}}{\partial x_{j\rangle }}
- p^{tot} \frac{\partial U_i}{\partial x_i},
\end{eqnarray}
where the constitutive relations \eqref{eq:const_stress_total} have been used in order to simplify the derivation. Therefore three further constitutive equations 
%\begin{widetext}
\begin{eqnarray}
 \hat{c}_{ij} q_j % \langle M^\prime_j  M^\prime_k  M^\prime_k  \rangle
           +\frac{1}{2}\rho \hat{\gamma}_i    \left( u^{(4)} 
           			   -   (u^{(2)}) ^2\right)  
           			   & =&\frac{3}{5}nbYq_i - wc_v\frac{\partial T}{\partial x_i} 
           			    - \frac{1}{2}\rho \hat{\Lambda}  \left(u^{(4)}_i
           			    - u^{(2)}_i  u^{(2)}  \right )
\label{eq:const_heat_total}
\end{eqnarray}
%\end{widetext}
have to be fulfilled in accordance with the energy evolution resulting from the Enskog equation i.e. Eq.~\eqref{eq:heat}.
\end{enumerate}
By constructing the constitutive system \eqref{eq:const_stress_total} and \eqref{eq:const_heat_total}, 12 equations are obtained for the nine unknowns of the tensor $\hat{\bm c}$ together with three unknowns of $\hat{\bm \gamma}$. \\ \ \\
\noindent {\color{myred}Let us} summarize the introduced closures in order to have a more clear form of the cubic dense FP model (henceforth referred as DFP). Therefore, the governing equation of the DFP model takes the form
\begin{eqnarray}
\label{eq:DFP}
\frac{\partial \mathcal{F}}{\partial t} 
+ \frac{ \partial }{\partial x_i}(V_i \mathcal{F}  ) 
&=& - \frac{ \partial}{\partial V_i} (A_i \mathcal{F}  )
+\frac{\partial^2}{\partial V_i \partial V_i} 
       \left( \frac{kTY}{m \tau} \mathcal{F}\right )- \frac{ \partial}{\partial x_i} (\hat{A}_i \mathcal{F}  ),
\end{eqnarray}
where $\bm A$ and $\hat{\bm A}$ read the expansions \eqref{eq:cubic_ansatz_ideal_gas}
 and \eqref{eq:2nd_A}, respectively. The nine coefficients $\bm c$ and $\bm \gamma$ come from the constitutive relations \eqref{eq:sys_cubic_FP_1}-\eqref{eq:sys_cubic_FP_2}, whereas the 12 coefficient $\hat{\bm c}$ and $\hat{\bm \gamma}$ follow Eqs.~\eqref{eq:const_stress_total} and \eqref{eq:const_heat_total}. Furthermore, the time scale $\tau=2\mu^{kin}/p$. Finally, the stability coefficients $\Lambda$ and $\hat{\Lambda}$ are obtained from Eqs.\eqref{eq:lambda} and \eqref{eq:Lambda*}, respectively.
 
 %%%%%%%%%    Here is one paragraph on limitations of DFP
 {\color{myred}
 \subsubsection{Accuracy limitations}
 \noindent In the derivation of the DFP model, three main approximations are involved which limit the accuracy of the model. In the following we provide rough account of the confidence region of the DFP model in terms of Kn, $nb$ and Mach number.
 \begin{enumerate}
 \item {\it Accuracy with respect to Kn:} Similar to the cubic FP approximation of the {\color{myred}Boltzmann equation}, here only moments up to the heat fluxes are matched with respect to the Enskog operator. Since moments higher than heat fluxes are not modeled accurately, a departure between the DFP model and the Enskog equation is expected for very large Knudsen numbers. From a body of conducted validation studies, it has been found that the cubic FP model gives rise to accurate results for Knudsen numbers up to around 2 \cite{Gorji2011,Gorji2015a}. Therefore we expect the same range of validity for the DFP model and hence $\textrm{Kn} \le 1$ should be respected for the application of the model. 
 \item {\it Accuracy with respect to nb:} The second source of errors introduced in the derivation of the DFP model lies on the approximations involved in Eq.~\eqref{eq:psi_phi} besides the closure of Y i.e. Eq.~\eqref{eq:Y}. Note that unlike the Boltzmann production terms, simplifications are necessary in order to obtain closed form expressions for the moments of the Enskog operator. Due to the small value of $\sigma$, the Taylor expansion of functions evaluated at $(\bm x\pm \sigma k)$ and $(\bm x\pm \sigma k/2)$ with respect to $\bm x$, becomes relevant. However the truncation of higher order terms in the expansion \eqref{eq:psi_phi} comes with the price of an error of order $nb$. Thus an upper bound of the form $nb\le 1$ is necessary for application of the DFP model.    
 \item {\it Accuracy with respect to Ma:} The last source of limitations comes from the molecular model employed in the derivation of the moments of the collision operator. So far we have only considered Maxwell interaction law for evaluation of the decay rates appearing in Eqs.~\eqref{eq:relaxations_stress_enskog}-\eqref{eq:relaxations_heat_enskog}. While it can be shown that the corresponding Maxwell terms are first order approximations of the more realistic hard sphere model~\cite{gupta2012automated}, it is well known that the shock thickness is sensitive to the molecular interaction law~\cite{Bird}. Therefore the DFP model in its current form may not lead to accurate shock profiles in comparison to the hard sphere or variable hard sphere molecular potentials. In the follow up study, we plan to further improve the DFP model appropriate for more realistic potentials.  
 \end{enumerate}
 }

\section{\label{sec:solution_algorithm}Solution Algorithm}
\noindent In this section, first the introduced DFP model is translated into a set of stochastic processes. Afterwards, a time integration scheme is derived for the resulting SDEs, where energy conservation is honored. Next, the ensemble averaging is introduced for evaluation of the statistical moments. At the end, an outline of the particle algorithm using the DFP model is presented. %Note that due to the presence of spatial derivatives, special care is taken such that the derivatives are computed without introducing numerical stencils.  

%%%%%%%%%%%%%%%%%%%%%%%%%%%%%%%%%%%%%%%%%%%%%%%%%%%%
%%%%%%%%%%%%%%%%%%%%%%%%%%%%%%%%%%%%%%%%%%%%%%%%%%%%
%%%%%%%%%%%%%%%%%%%%%%%%%%%%%%%%%%%%%%%%%%%%%%%%%%%%
\subsection{It{\={o}} Process}
\noindent Due to the high dimensionality of the phase space, the stochastic representation of the governing equation \eqref{eq:DFP} is considered. Suppose $\bm M(t)$ and $\bm X(t)$ are random variables associated with the sample spaces $\bm V$ and $\bm x$, respectively. By applying the It{\={o}} lemma, Eq.~\eqref{eq:DFP} can be translated to 
\begin{eqnarray}
dM_i&=&A_idt+\sqrt{\frac{2kT Y}{m\tau}}dW_i \label{eq:ito-vel} \\
\textrm{and} \ \ \ \ dX_i&=&\tilde{A}_idt+M_idt, \label{eq:ito-pos}
\end{eqnarray}
where $d\bm W$ is the Wiener increment with $\langle d W_i \rangle =0$ and $\langle dW_idW_j \rangle=\delta_{ij}dt$. Here $\langle ...\rangle$ denotes the expectation and $\delta_{ij}$ is Kronecker delta.
Equivalently, many realizations of the system~\eqref{eq:ito-vel}-\eqref{eq:ito-pos}  i.e. $\bm M^{(i)}$ and $\bm X^{(i)}$ with $i\in\{1,...,N\}$,  provide a density 
\begin{eqnarray}
\label{eq:f-delta}
f(\bm V,\bm x;t)&=&\lim_{N\to \infty}\frac{1}{N}\sum_{i=1}^N\delta \left(\bm M^{(i)}(t)-\bm V\right)\delta\left(\bm X^{(i)}(t)-\bm x\right),
\end{eqnarray}
which is related to the solution of the DFP Eq.~\eqref{eq:DFP} through $\mathcal{F}(\bm V,\bm x,t)=\mathbb{M} f(\bm V,\bm x;t)$ with $\mathbb{M}$ being the total physical mass of the system (see {\color{myred}Ref.}~\cite{Gorji2014a}).

\subsection{\label{sec:time_integration}Time Integration}
\noindent A time marching scheme is required in order to update the state of particle realizations which evolve based on the system~\eqref{eq:ito-vel}-\eqref{eq:ito-pos}. 
Since the structure of the velocity evolution i.e. Eq.~\eqref{eq:ito-vel} is similar to the ideal gas case, the time integration proposed in {\color{myred}Ref.}~\cite{Gorji2014} can be directly employed. Let variables with superscript $^{n}$ denote the approximate solution at time $t^n$ resulting from the time integration scheme. Therefore the update of the velocity at $t^{n+1}=t^{n}+\Delta t$ takes the form
%\begin{widetext}
\begin{eqnarray}
M_i^{n+1}&=& U_i^n 
+ \alpha_{cell}\tilde{M}^{n}_i,
\end{eqnarray}
where
\begin{eqnarray}
\tilde{M}^{n}_i&=& \Big (
 M_i^{\prime n} e^{-\Delta tY^n/\tau^n} + c^n_{ik} M_k^{\prime n} 
+ \gamma^n_i ( M_k^{\prime n} M_k^{\prime n} - 3kT^{n}/m ) 
\nonumber \\
&+&    \Lambda ( M^{\prime n}_i M^{\prime n}_j M^{\prime n}_j -  2q^{n}_i/\rho^{n})
+ \sqrt{\frac{kT^{n}Y^{n}}{\tau^{n} m}\left({1-e^{-2\Delta tY^{n} /\tau^{n}}}\right)} \xi_i
\Big ). \label{eq:FP_dense_velocity_update} 
\end{eqnarray}
Here $\xi_i$ is a normally distributed random number and  $\alpha_{cell}$ is the correction factor for kinetic energy
\begin{eqnarray}
\alpha_{cell} = 
\frac{\langle M_{j}^{\prime n} M_{j}^{\prime n} \rangle}
{\langle \tilde{M}_i^{n} \tilde{M}_{i}^{n} \rangle}~.
\label{alpha_cell}
\end{eqnarray}
Note that the fluctuating part of the velocity is $\bm M^\prime=\bm M-\bm U$. \\ \ \\
The position update follows the time integration of the SDE \eqref{eq:ito-pos}. For simplicity, the Euler approximation 
\begin{eqnarray}
X_i^{n+1} 
&=& X_i^{n} + M_i^{n}\Delta t + \Big ( \hat{c}^{n}_{ij} M^{\prime n}_j + \hat{\gamma}^{n}_i (M^{\prime n}_j M^{\prime n}_j - {3kT^n}/{m}) 
\nonumber \\
&+& \hat{\Lambda}^n ( M^{\prime n}_i M^{\prime n}_j M^{\prime n}_j -  2q^{n}_i/\rho^{n} \Big ) \Delta t  \label{eq:FP_dense_position_update}
\end{eqnarray}
is employed. \\ \ \\
Note that since the macroscopic coefficients involved in the drift and diffusion are treated as frozen coefficients in the numerical schemes \eqref{eq:FP_dense_velocity_update} and \eqref{eq:FP_dense_position_update}, the time step size should honor a {\color{myred}Courant-Friedrichs-Lewy} (CFL) type criterion. Therefore $\Delta t$ should be chosen such that on average, a particle crosses not more than one computational cell per time step (see {\color{myred}Ref.}~\cite{pfeiffer2017adaptive}). %Here, CFL condition is defined such that
%\begin{eqnarray}
%\Delta t \leq \frac{1}{2} \dfrac{h}{\theta}
  %\label{eq:CFL}
%\end{eqnarray}
%where $h$ denotes cell size and $\theta = \sqrt{kT/m}$ is thermal velocity.
%\end{widetext}

\subsection{\label{moment_evaluation}Moment Estimation }
\noindent The update scheme presented in Section \ref{sec:time_integration} requires estimations of different velocity moments. The identity \eqref{eq:f-delta} provides a means to a direct evaluation of the moments through arithmetic averages. Suppose $N_p$ computational particles indexed by the superscript $i$, each with the statistical weight $w$. Note that $w$ is related to the total mass via $\mathbb{M}=wN_p$. Furthermore, we consider a discretization of the physical domain into computational cells $\Omega^{(j)}$ with volumes $\delta \Omega^{(j)}$.  Moreover, if we assume that each computational cell is spatially homogeneous, the moment of an arbitrary velocity function $\bm Q(\bm V)$  can be estimated
\begin{eqnarray}
\int_{\mathbb{R}^3}\bm Q(\bm V)\mathcal{F}(\bm V,\bm x,t)d^3\bm V&\approx&\frac{1}{\delta \Omega^{(j)}}\sum_{i}w\bm Q(\bm M^{(i)}) \ \ \ ; \ \ \ \bm X^{(i)} \&\  \bm x \in \Omega^{(j)}.
\label{eq:ensemble_center}
\end{eqnarray}
While the above estimator is convenient for moment evaluations at cell centers, it does not provide an accurate estimate on cell faces. This can be crucial here since spatial gradients have to be found for closure systems \eqref{eq:const_stress_total} and \eqref{eq:const_heat_total}. 
Consider a cell face $\Gamma_{\Omega^{(j)}}$ with the area $\delta \Gamma_{\Omega^{(j)}}$ at one boundary of the cell $\Omega^{(j)}$. The moments evaluated on $\Gamma_{\Omega^{(j)}}$ can be found through the particles that crossed the face during the time step $\Delta t$. However the statistical weight of particles should be adjusted for the time that they spent in the infinitesimal volume around $\Gamma_{\Omega^{(j)}}$. Therefore the resulting expression for the moment evaluation on the face becomes 
%considered as an infinitesimal volume, therefore the time spent of particle in this thin layer we are able to calculate the quantity  $\bm{Q}(\bm{v})$ on the faces of cell.
%
\begin{eqnarray}
\int_{\mathbb{R}^3}\bm Q(\bm V)\mathcal{F}(\bm V,\bm x,t)d^3\bm V&\approx&\frac{1}{\Delta t\ \delta \Gamma_{\Omega^{(j)}}}\sum_{i}\frac{w\bm Q(\bm M^{(i)})}{|\bm M^{(i)}\cdot \bm n_{\Gamma_{\Omega^{(j)}}}|} \ \ \ , \ \ \   \bm x \in \Gamma_{\Omega^{(j)}},
\label{eq:ensemble_face}
\end{eqnarray}
for all particles that crossed the face $\Gamma_{\Omega^{(j)}}$ during $\Delta t$. Here $\bm n_{\Gamma_{\Omega^{(j)}}}$ is the normal of the cell face. The cell face moment estimator i.e. Eq.~\eqref{eq:ensemble_face} provides a convenient way to calculate gradients of a macroscopic quantity. By using a finite volume approximation, spatial gradients can be rewritten as surface integrals and thus reduce to a function of the cell face averages.

\subsection{Particle Algorithm}
\noindent In order to setup a particle simulation for the DFP model, first particles should be generated according to an initial condition $\mathcal{F}_0$. The statistical weight $w$ should be chosen such that a reasonable number of particles will be required for simulations. Furthermore a spatial discretization of the physical domain is adopted in order to sample different macroscopic quantities. Once particles are initialized, the steps outlined in Table~\ref{tab:DFP_outline} are followed until the final time is reached. In the case of statistically stationary simulations, the algorithm is iterated until an appropriate level of the statistical error is obtained through time averaging of the sampled moments. Note that treatments of inflow/outflow or wall boundary conditions are similar to DSMC (see e.g. {\color{myred}Ref.}~\cite{Bird}).
%\noindent Algorithm of proposed cubic  Fokker-Planck model for dense gases (DFP)  is described in Table~\ref{tab:DFP_outline}.
\begin{table}[]
\centering
\caption{Outline of the DFP solution algorithm during each time step.}
\begin{tabular}{|ll|}
\hline
 1. &  Calculate velocity moments at cell centers by Eq.~\eqref{eq:ensemble_center} and cell faces by Eq.~\eqref{eq:ensemble_face} .\\
 2. &  Calculate $c_{ij}$ and $\gamma_i$ from Eqs.~\eqref{eq:sys_cubic_FP_1}-\eqref{eq:sys_cubic_FP_2}, and $\hat{c}_{ij}$ and $\hat{\gamma}_i$ from Eqs.\eqref{eq:const_stress_total} and \eqref{eq:const_heat_total} . \\
% 3. &  Calculate $\alpha_{ij}$ and $\beta_i$ from Eq.~(\ref{eq:2_shear_stress_correction}) and Eq.~(\ref{eq:heat_flux_expression}). \\
 3. &  Update the velocity according to Eq.~(\ref{eq:FP_dense_velocity_update}).  \\
 4. &  Stream particles using Eq.~(\ref{eq:FP_dense_position_update}).  \\
 5. &  Apply boundary conditions.  \\
 6. &  Sample data.\\
 \hline  
\end{tabular}
\label{tab:DFP_outline}
\end{table}
%%%%%%%%%%%%%%%%%%%%%%%%%%%%%%%%%%%%%%%%%%%%
%%%%%%%%%%%%%%%%%%%%%%%%%%%%%%%%%%%%%%%%%%%%
%%%%%%%%%%%%%%%%%%%%%%%%%%%%%%%%%%%%%%%%%%%%
\section{\label{sec:results} Results}
\noindent For assessment of the accuracy and performance of the derived DFP model, simulations of various canonical test cases were carried out. The results are compared with respect to ESMC simulations, serving as the benchmark. First, deviation of the equilibrium pressure from the ideal gas law is tackled by the DFP simulation of a gas confined in an insulated box. Next, the planar Couette flow is considered, where the shear stress generated by the moving walls is investigated. Then, the planar Fourier flow is simulated by imposing different temperatures at the walls.
{\color{myred}At last, a lid-driven cavity flow is studied by moving only one of the walls.}
In all simulations, the hard sphere model of argon was employed with $m=6.6335 \times 10^{-26}$~kg and $\mu^{\textrm{kin}}=2.117 \times 10^{-5}$~$\textrm{kg}/(\textrm{m}\textrm{s})$ at $T=273$~K. 
\subsection{\label{sec:box}Equilibrium Box}
\begin{figure}
  \centering
  \begin{subfigure}[b]{0.48\columnwidth}
  \scalebox{1}{\includegraphics[ angle =0]{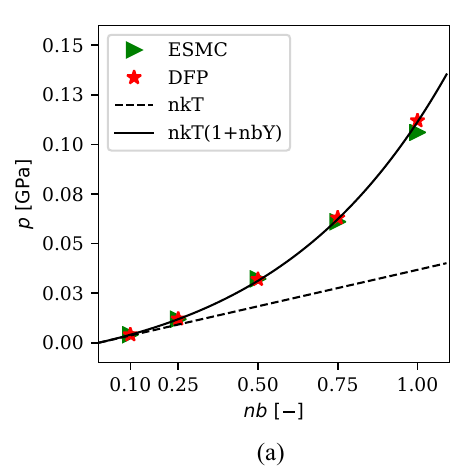}}
    %\caption{}
  \end{subfigure}
  \
  \begin{subfigure}[b]{0.48\columnwidth}
   \scalebox{1}{\includegraphics[ angle =0]{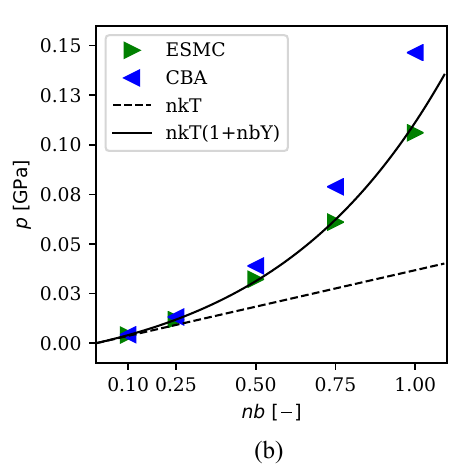}}
    %\caption{}
  \end{subfigure}
  \caption{Equilibrium pressure of gas obtained from  {\color{myred} DFP and CBA models in comparison with ESMC, shown in (a) and (b), respectively.} Solid black line indicates $p=n k T(1+nbY)$ and dashed line is $p=nkT$.}
  \label{fig:P}
\end{figure}

\noindent In order to check whether the proposed DFP model gives us the pressure consistent with the Enskog solution, equilibrium conditions are imposed on a gas confined in a box. The gas is at rest with no external force and no temperature gradient. Thus, the macroscopic quantities of the gas are expected to remain constant in time and the VDF stays Maxwellian.
Initially, positions of particles are uniformly distributed inside the box where velocities are set based on the Maxwellian distribution at temperature of $273$~K. The walls are insulated and thus specular reflection is considered as the boundary condition. The box side length is chosen such that $\textrm{Kn}=0.01$ is achieved. 
% Walls are thermal walls with no velocity.
%The iso-thermal stationary walls at $T_w=273 \ K$ are considered.
  % From that point on, simulation results are read for visualization. 
The pressure is calculated from the average force {\color{myred}per unit area} that wall experiences from gas molecules, i.e.
\begin{flalign}
p = \frac{m}{A\ t^f} \sum_{j=1}^{N_w} ( \bm M^{\prime,(j)}-\bm M^{(j)})\cdot \bm n ,
\end{flalign}
where $A$ is the face area, $(\bm M^{(j)},\bm M^{\prime,(j)})$ are the incoming and reflecting velocity of the particle $j$, respectively and $\bm n$ normal of the wall. The final averaging time is set to $t^f$ and $N_w$ indicates the number of particles that hit the wall up to $t=t^f$. \\ \ \\
\noindent The box is uniformly discretized into 100 computational cells and $1000$ particles per cell were employed. Furthermore, the time step size for DFP, CBA and ESMC are set based on the mean collision time and CFL condition to $\Delta t \in \{ 33.9,12.3,5., 2.87,1.78\}\times 10^{-14}$~s for $nb \in \{ 0.1, 0.25, 0.75, 1.0\}$ accordingly. 
\noindent The obtained pressure results are shown in Fig.~\ref{fig:P}, where the accuracy of the proposed DFP model is compared with respect to ESMC and CBA. It is interesting to see that the proposed DFP model is in a better agreement with ESMC in comparison to CBA, where the latter over-predicts the value of $nkT(1+nbY)$.

% are in obtaining  pressure at equilibrium compared to CBA. DSMC and FP of ideal gas predict equation of state for ideal gas $P=nk_bT$ while dense FP and CBA try to capture $P=nk_bT(1+nbY)$. Temperature of walls are set to $T_w = 273 \ K$ and walls are kept constant $U_w=0$. Proposed FP indeed gives a better approximation to $P=nk_bT(1+nbY)$ since CBA over predicts pressure.

\subsection{\label{sec:couette_flow}Couette Flow}

\begin{figure*}[t]
    \centering
        \centering
          \begin{subfigure}[b]{0.48\columnwidth}
          \includegraphics{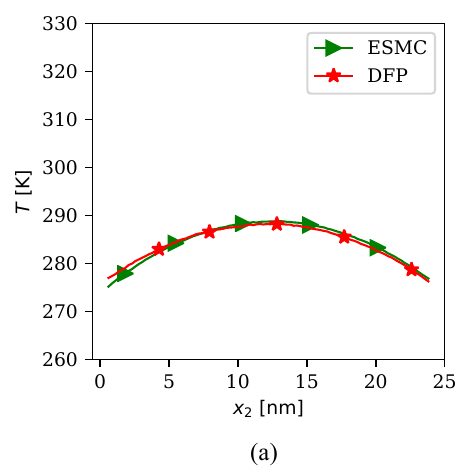}
%    \caption{}
  \end{subfigure}
   \hfill
        \centering
        \begin{subfigure}[b]{0.48\columnwidth}
        \includegraphics{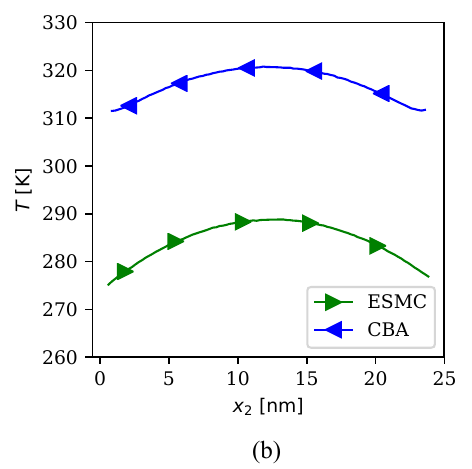}
   %     \caption{}
          \end{subfigure}
           \hfill
        \centering
        \begin{subfigure}[b]{0.48\columnwidth}
                \includegraphics{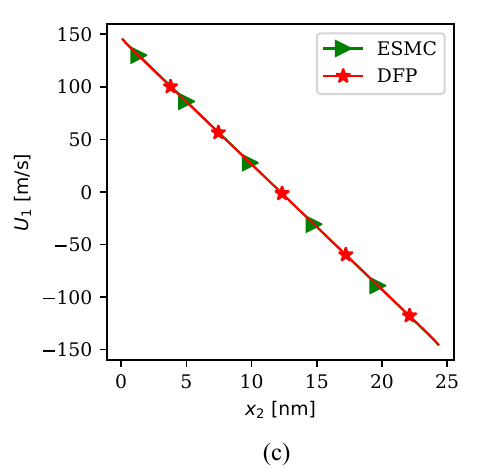}
           %             \caption{}
          \end{subfigure}
          \hfill
        \centering
        \begin{subfigure}[b]{0.48\columnwidth}
        \includegraphics{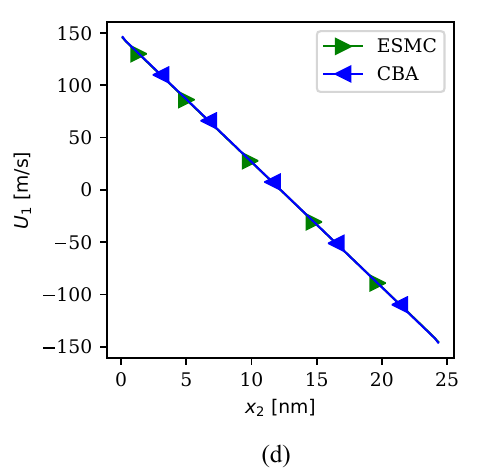}
        % \caption{}
          \end{subfigure}
    \caption{
   {\color{myred} Couette flow results using DFP, CBA and ESMC simulations for $\textrm{Kn}=0.01$ and $nb=0.5$. Temperature profiles are shown in (a) and (b) for DFP and CBA, respectively. The velocity profiles are shown in (c) and (d) for DFP and CBA, respectively.}}
    \label{fig:couette_flow_U_T}
\end{figure*}

\begin{figure}
\centering
\begin{subfigure}[b]{0.48\columnwidth}
\scalebox{1}{\includegraphics[ angle =0, clip=true, trim =0mm 0mm 0mm 0mm]{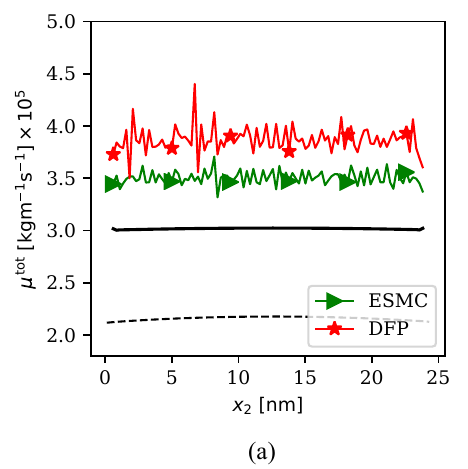}}
%\caption{}
\end{subfigure}
\begin{subfigure}[b]{0.48\columnwidth}
\scalebox{1}{\includegraphics[ angle =0, clip=true, trim = 0mm 0mm 0mm 0mm]{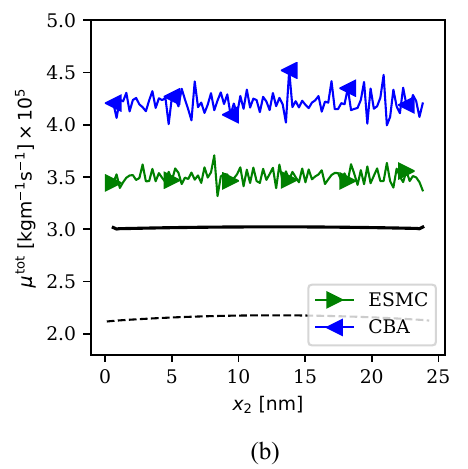}}
%\caption{}
\end{subfigure}
\caption{The total viscosity from Couette flow simulations at $\textrm{Kn}=0.01$ and $nb=0.5$ using {\color{myred}DFP and CBA models in comparison with ESMC, depicted in (a) and (b), respectively.} Solid and dashed lines indicate Chapman-Enskog expressions of viscosity for the dense and dilute limit, respectively.}
\label{fig:mu_couette}
\end{figure}

\noindent To further examine the accuracy of the DFP model in a shear dominated flow, the planar Couette flow is considered. 
% As explained in Section \ref{sec:Collisional_Transfer_of_Molecular_Quantities}, one can relate the pressure tensor to the gradients of the bulk velocity.
{\color{myred}Let us} consider a planar Couette setup with $x_2$-direction being normal to the walls. %the walls of the box be large enough in $x_1$ and $x_3$ dimensions. Thus,  one can assume that the gas behavior is one dimensional in $x_2$ dimension. 
The walls have velocity of  {\color{myred}$U_w = (\pm 150, 0, 0)^T \ \mathrm{ms}^{-1}$} and the distance $L$ between them is adopted such that $\textrm{Kn}=0.01$ is achieved. The chosen number density results in $nb=0.5$. Convergence studies on the spatial refinement lead us to $100$ computational cells in $x_2$.
Besides, in average $1000$ particle per cell were employed. Furthermore, the time step size $\Delta t =5.125 \times 10^{-14}$~s is set for all CBA, ESMC and DFP simulations. Around 20,000 number of time steps were used to reach statistically stationary condition and afterwards sufficient number of time steps for averaging was adopted to control the statistical errors.  \\ \ \\
The results characterizing the flow field are shown in Fig.~\ref{fig:couette_flow_U_T}, where the mean velocity and temperature are depicted for DFP, ESMC and CBA models.  Note that while the velocity profiles are identical for all three models, the CBA model reveals a significant overshoot for the temperature with respect to ESMC and DFP.
\\ \ \\ \noindent A closer inspection of the models can be achieved by measuring the equivalent viscosity of the gas and comparing it with respect to the Chapman-Enskog prediction. %Next, the exerted particle force on each wall is evaluated from ESMC, CBA and DFP simulations. %Dimensions and number density are set to reach $Kn=0.01$ and $nb=0.5$.
%\subsubsection{Viscosity}
% 1D corresponds only to position of particles meaning velocity still has three non-zero components. 
%\begin{enumerate}
%\item{{\it Chapmann-Enskog}:} 
\noindent In Chapman-Enskog's limit of small Kn, by ignoring derivatives with respect to $x_1$ and $x_3$ one can show that the shear stress for %ideal and 
 dense gases reads
\begin{flalign}
%p^{\text{kin}}_{ 12 } &= 
%-   \mu^{\mathrm{kin}} \frac{\partial U_1}{\partial x_2} \label{eq:pkin_chapmann-enskog}\\
 %\text{and} \ \ \ \ \ \ 
 \pi^{\text{tot}}_{12} &= 
-\Big( \frac{\mu^{\mathrm{kin}}}{Y} ( 1+2nbY/5)^2  + \frac{3w}{5} \Big) \frac{\partial U_1}{\partial x_2}~.
\label{eq:ptotal_chapmann-enskog}
\end{flalign}
%
%respectively.
%The dilute  gas regime should give us $\mu^{kin}$
% as $p^{ideal}_{\langle 12 \rangle}/({\partial U_1}/{\partial x_2})$ while models for dense gas should provide 
Therefore the viscosity of the dense gas based on the Chapman-Enskog expansion becomes
 $\mu^{\text{tot}} = {\mu^{\text{kin}}} ( 1+2nbY/5)^2/Y+{3w}/{5}$.  \\ \ \\
  %as $p^{dense}_{\langle 12 \rangle}/({\partial U_1}/{\partial x_2})$.
%\item{{\it Particle simulations}:} 
\noindent Alternatively, the total shear stress can be estimated from particles through Eq.~\eqref{eq:pressure_tensor_enskog} which reduces to
\begin{eqnarray}
\pi^{\text{tot}}_{12}  &=&  ( 1+2nbY/5) \pi_{12}
- (3w/5)  \frac{\partial U_{1}}{ \partial x_{2 }}
\end{eqnarray}
for the planar setup considered here. Therefore by dividing the obtained $\pi^{\text{tot}}_{12}$ by the velocity gradient $\partial U_1/\partial x_2$, an equivalent viscosity can be calculated from particle simulations. \\ \ \\
%\end{enumerate}
 %Therefore, in order to compare the calculated viscosities at the reference temperature $T_0 = 273 \ K$,  $\mu(T)$ has to be scaled back to the reference temperature.
\noindent Figure~\ref{fig:mu_couette} depicts the measured viscosity from three different simulations using ESMC, CBA and DFP models. %for each model. 
For the reference, the Chapman-Enskog limits are also plotted. Since the viscosity varies with respect to the temperature and number density, the viscosity values calculated from the Chapman-Enskog expansion are evaluated with the temperature and number density resulting from the ESMC simulation. Overall, the agreements are close between all models, yet CBA and DFP over predict the total viscosity with respect to the ESMC result.  
%
% i.e. $\mu^{\text{kin}}_0 = 2.117 \times 10^{-5} [\mathrm{kg\ m^{-1}s^{-1}}]$ as in black dashed line and $\mu^{\text{tot}}_0$ with bulk effects as black solid line. Subscript $0$ denotes the reference values set for the flow. As expected, FP of ideal gas and DSMC predict $\mu^{\text{kin}}_{0}$ with a decent accuracy.
%However, ESMC does not converge to limit of Chapmann-Enskog expansion in such a high density and velocity gradient. 
%The error is assumed to be associated with the the first approximation committed by assuming Eq.~(\ref{eq:pressure_tensor_enskog}) and Eq.~(\ref{eq:1st_approximation}).
%Proposed DFP commits error compared to  ESMC which needs to be investigated. Measured viscosity from CBA simulation is even more erroneous than DFP when it is compared to the benchmark. 

\subsection{\label{fourier_flow}Fourier Flow}
\begin{figure}
\centering
\begin{subfigure}[b]{0.48\columnwidth}
\scalebox{1}{\includegraphics[ angle =0, clip=true, trim = 0mm 0mm 0mm 0mm]{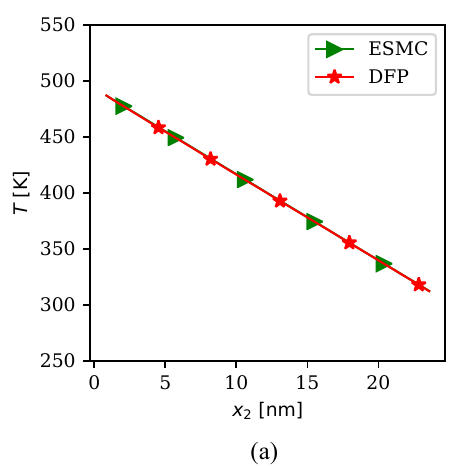}}
%\caption{}
\end{subfigure}
\ \ \ 
\begin{subfigure}[b]{0.48\columnwidth}
\scalebox{1}{\includegraphics[ angle =0, clip=true, trim = 0mm 0mm 0mm 0mm]{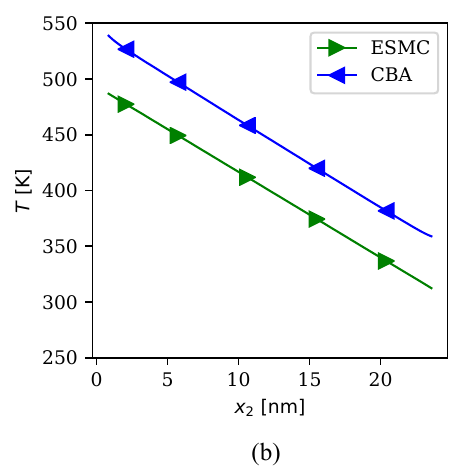}}
%\caption{}
\end{subfigure}
\caption{ Temperature profiles for the Fourier flow with $\textrm{Kn}=0.01$ and $nb=0.5$ using  {\color{myred} DFP and CBA models in comparison with ESMC, shown in (a) and (b), respectively}.}
\label{fig:T_fourier}
\end{figure}

\begin{figure}[t]
\centering
\begin{subfigure}[b]{0.48\columnwidth}
\scalebox{1}{\includegraphics[ angle =0, clip=true, trim = 0mm 0mm 0mm 0mm]{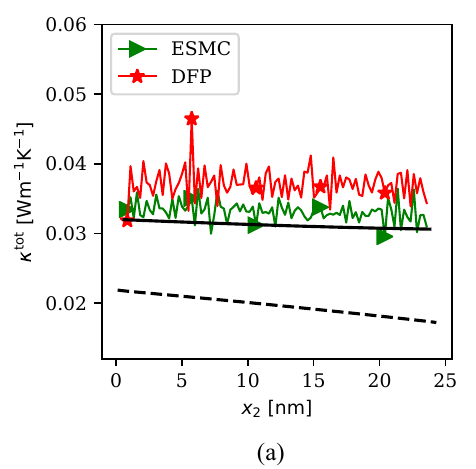}}
%\caption{}
\end{subfigure}
\ \ \ 
\begin{subfigure}[b]{0.48\columnwidth}
\scalebox{1}{\includegraphics[ angle =0, clip=true, trim = 0mm 0mm 0mm 0mm]{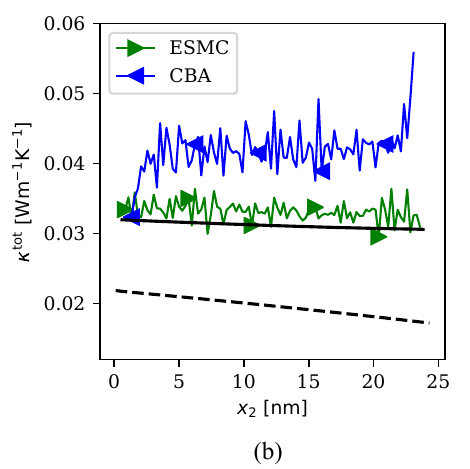}}
%\caption{}
\end{subfigure}
\caption{
The heat conductivity coefficient calculated from the Fourier flow with $\textrm{Kn}=0.01$ and $nb=0.5$ using {\color{myred}DFP and CBA models in comparison with ESMC, depicted in (a) and (b), respectively.} Solid line indicates $\kappa^{\text{tot}} = \kappa^{\text{kin}} (1+3 n b Y/5)^2/Y+w c_v$ and dashed line  $\kappa^{\text{kin}}$.}
\label{fig:k_fourier}
\end{figure}

\noindent For further assessment of the DFP model, here we investigate the Fourier flow where heat transport phenomena play the major role.
%In this section, the heat conduction of dense will be evaluated in the dense regime. As explained in Section \ref{sec:cubic_FP}, the cubic model is proposed to enforce the dense effect on the heat conduction of gas.
The setup is similar to the one of the Couette flow in Section~\ref{sec:couette_flow}, except for the boundary conditions. Here, by imposing a temperature difference on the walls,  the gas expands along the temperature gradient. The temperature of the fully diffusive walls are set to $T_{w1} = 300$~K and $T_{w2} = 500$~K. The initial number density is chosen such that $nb=0.5$. The distance between the two walls comes from $\textrm{Kn}=0.01$. Similar to the Couette flow simulations, it is observed that $100$ computational cells provide a fine enough accuracy. Accordingly $\Delta t=3.787 \times 10^{-14}$~s is employed for the DFP model, CBA and ESMC simulations. \\ \ \\
The temperature profiles resulting from simulations based on ESMC, CBA and DFP models are shown in Fig.~\ref{fig:T_fourier}. While good agreement between ESMC and DFP results can be observed, the CBA approach gives rise to a temperature overshoot; similar to the Couette flow scenario. 
\\ \ \\
% are more than enough 
% for capturing the temperature gradient. Also note that $1000$ particles per cell was employed in all the simulations. 
%Similarly, one can again expect  an one dimensional flow which is driven due to the temperature difference of the walls.
% Again, $\partial U_k/ \partial x_k \rightarrow 0$ due to symmetry.
 %Fourier's law together with the  Chapman-Enskog limit of heat conductivity  Eq.~(\ref{eq:enskog_heat_gradinet_T}) give us
 %, i.e. \cref{heat_conduction_enskog_nsf}, gives us
\noindent Similar to the viscosity evaluation performed for the Couette flow results, the heat conductivity can be extracted from the Fourier flow. The Chapman-Enskog expansion leads to
\begin{eqnarray}
%q_2^{\text{kin}} &=& - \kappa^{\text{kin}} \frac{\partial T}{\partial x_2} ~\text{and}
%\label{eq:fourier_x2}
 %\\
q_2^{\text{tot}} &=& - \Big ( \frac{\kappa^{\text{kin}}}{Y} (1+3nbY/5)^2 + c_v w \Big ) \frac{\partial T}{\partial x_2},
\label{eq:fourier_x2_tot}
\end{eqnarray}
 for the dense gas. Therefore, the heat conductivity resulting from the Chapman-Enskog limit reads $\kappa^{\text{tot}} = \kappa^{\text{kin}}(1+3nbY/5)^2/Y+ c_v w$. \\ \ \\%This can be evaluated  this transport coefficient  by calculating the heat flux $q_2$ and the temperature gradient $\partial T/ \partial x_2$ from simulations.
 %The temperature gradient is calculated based on finite volume assumption and temperature on the faces of cell according to the time spent by particles crossing the face, Section~\ref{moment_evaluation}.
 %using a central finite difference on the time averaged temperature profile.
% Heat flux however is a bit tricky. To avoid stochastic noise in heat flux calculation, it has been calculated again as a polynomial of averaged velocity moments.
 %The heat flux of ideal gas can be calculated  based on second moment of fluctuating velocity
% \begin{widetext}
% \begin{eqnarray}
%q_{2}^{kin} &=& \frac{1}{2} \rho \langle M_2^\prime  M_j^\prime M_j^\prime \rangle \nonumber \\
%&=&\frac{1}{2}  \rho \langle (M_2-U_2) ( M_j-U_j) ( M_j-U_j) \rangle \nonumber \\
%&=&\frac{1}{2}  \rho \Big( \langle M_2 M_j M_j \rangle -2\langle M_2 M_j \rangle U_j - \langle M_j M_j \rangle U_2  +2\langle M_j \rangle U_j U_2 \Big).
%\end{eqnarray}
%\end{widetext}
%In dense gas, Eq.~(\ref{eq:heat_conduction_enskog}) can be employed to calculate $q_2^{\text{tot}}$ using moment samples. 
\noindent Furthermore, the total heat flux
\begin{eqnarray}
q_2^{\text{tot}} &= &
(1+3nbY/5) q_2
- c_v w  \frac{\partial T}{\partial x_2}
%\label{eq:heat_conduction_enskog}
\end{eqnarray}
can be directly estimated from particles ensemble for the planar setup considered here. Through dividing the resulting heat flux by the temperature gradient, the heat conductivity coefficient is evaluated from particle simulations. \\ \ \\
Figure~\ref{fig:k_fourier} shows the thermal conductivity calculated from the Fourier flow using DFP, ESMC and CBA based simulations. The solid and dashed lines indicate the Chapman-Enskog expressions of the heat conductivity for dense and dilute gases, respectively. Note that the Chapman-Enskog values are evaluated at the temperature and number density predicted by the ESMC simulation. While certain discrepancies can be observed between the DFP and ESMC results, overall a good agreement is shown. Yet the CBA model leads to a larger error in the heat conductivity coefficient in comparison to the DFP model. 
%Similar to Section \ref{sec:couette_flow}, ESMC at  high densities such as this problem does not guarantee $\kappa^{tot}$.
%It follows from the fact that ESMC as exact solution of Enskog equation satisfies   Eq.~(\ref{eq:enskog_heat_gradinet_T}) admitting error in derivation. Since Eq.~(\ref{eq:heat_conduction_enskog}) was deployed to design DFP, heat flux is investigated furthermore.
{\color{myred}
\subsection{\label{lid_driven_cavity} Lid-Driven Cavity}
\noindent The higher level of closure implied by the kinetic models allows for capturing highly non-equilibrium phenomena that may not be accurately described using conventional hydrodynamics. The lid-driven cavity test case, is one example where cold to hot heat fluxes can be observed inside the gas using kinetic based simulations, in contrast to Fourier's law \cite{Gorji2014,mohammadzadeh2012thermal,rana2013robust,john2010investigation}. In order to examine the accuracy of the introduced DFP model in such a challenging scenario, the dense gas simulations of the lid driven cavity are conducted here using the DFP model and ESMC.  The initial gas density and geometric dimensions are set such that  $nb=0.1$ and $\textrm{Kn}=0.1$. The diffuse isothermal walls with temperature $T_w = 273 \ \textrm{K}$  are employed, where the upper wall is moved with velocity $\bm U_w = (300,0,0)^T \textrm{ms}^{-1}$. The physical domain is discretized by a $150\times150$ uniform grid and 
 $\Delta t=1.777 \times 10^{-13}$~s was employed according to the CFL condition. On average $1000$ particles per cell were utilized and sampling for post-processing was performed after reaching statistically stationary condition; roughly after $3000$ time steps.

\begin{figure}[t]
\centering
\begin{subfigure}[b]{0.48\columnwidth}
\scalebox{1}{\includegraphics[width=8.2cm, keepaspectratio=true,clip=true, trim = 0mm 1mm 1mm 0mm]{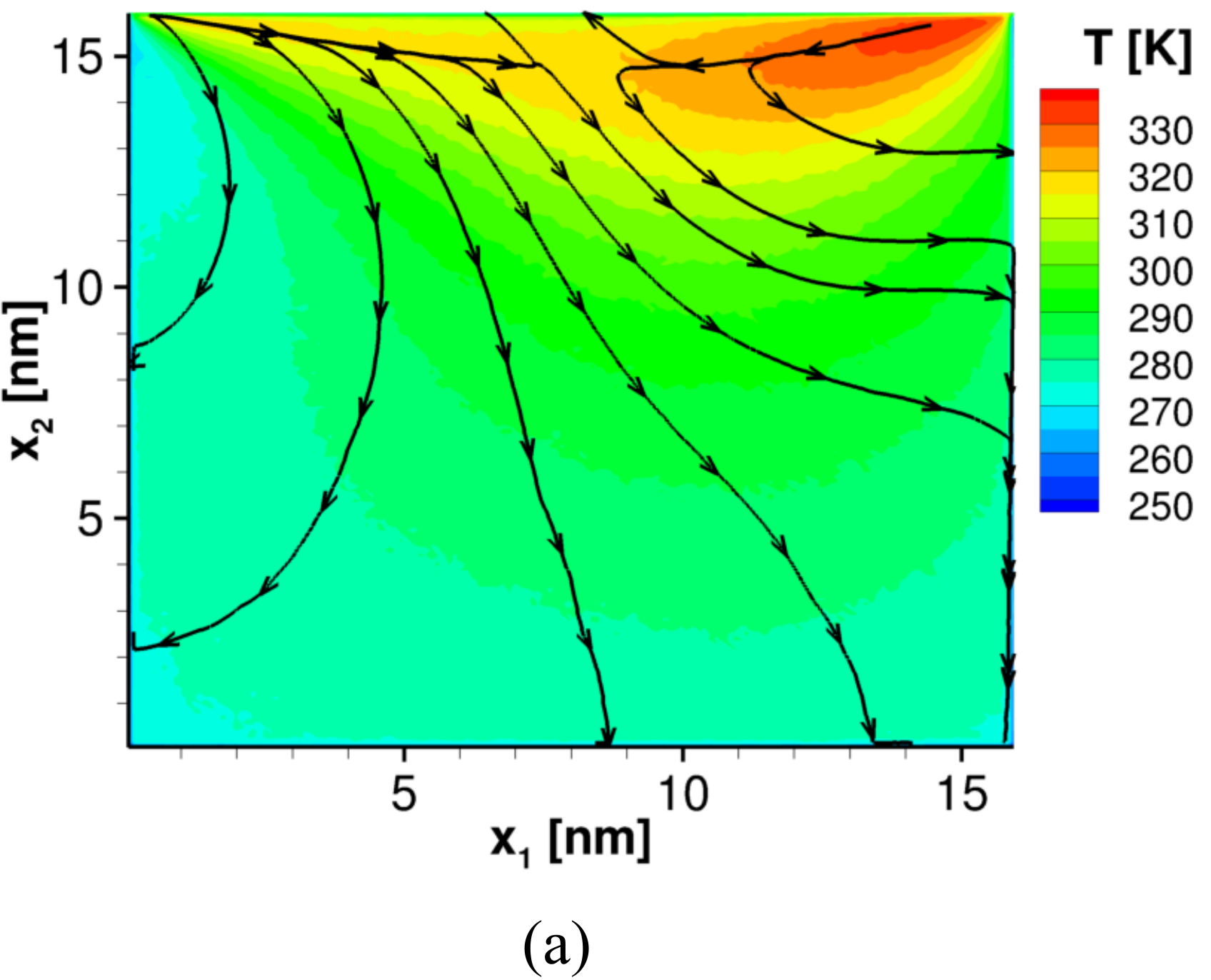}}
   % \caption{}
  \end{subfigure}
  \ \ \ 
  \begin{subfigure}[b]{0.48\columnwidth}
\scalebox{1}{\includegraphics[width=8.2cm, keepaspectratio=true,clip=true, trim = 0mm 1mm 1mm 0mm]{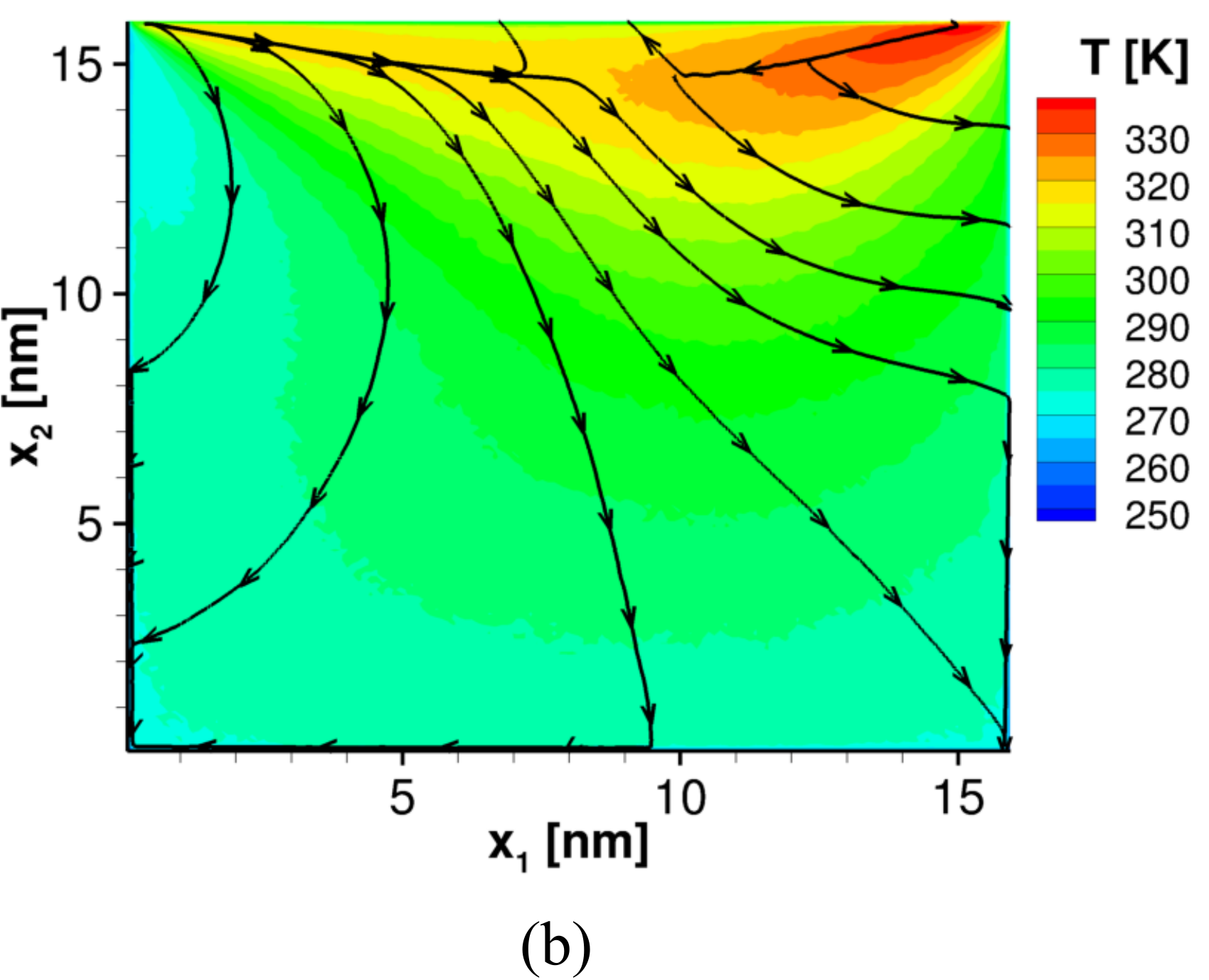}}
 %   \caption{}
  \end{subfigure}
\caption{
{\color{myred}Temperature contours overlaid by the integral heat flux curves in the lid-driven cavity flow using (a) the DFP model and (b) ESMC.}}
\label{fig:lid_driven_contour}
\end{figure}
\noindent As shown in Figure~\ref{fig:lid_driven_contour},  both DFP and ESMC simulations admit the counter Fourier heat fluxes. Consistency between DFP and ESMC can be furthermore investigated by analyzing the heat flux value across the domain.% a horizontal line on upper part of domain. 
The total heat fluxes $q_1^{\text{tot}}$ and $q_2^{\text{tot}}$ are depicted along the line $x_2 = 4L/5$, in
Figure~\ref{fig:lid_driven_over_line}. Very good agreement is observed between DFP and ESMC results.

\begin{figure}[t]
\centering
\begin{subfigure}[b]{0.49\columnwidth}
\scalebox{1}{\includegraphics[clip=true, trim = 0mm 0mm 0mm 0mm]{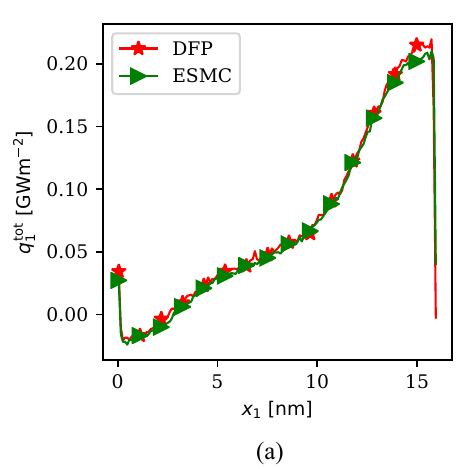}}
   % \caption{}
  \end{subfigure}
  \
  \begin{subfigure}[b]{0.49\columnwidth}
\scalebox{1}{\includegraphics[clip=true, trim = 0mm 0mm 0mm 0mm]{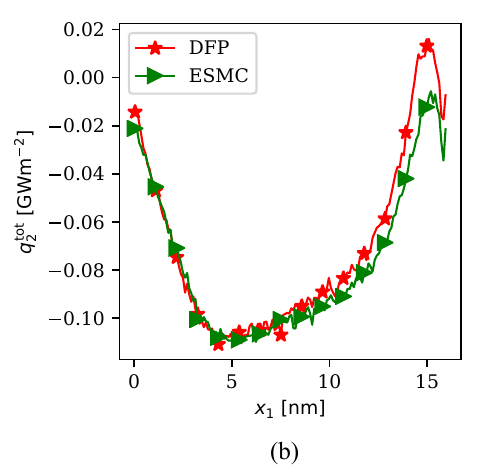}}
  %  \caption{}
  \end{subfigure}
\caption{{\color{myred}Total heat fluxes (a) $q_1^{\text{tot}}$ and (b) $q_2^{\text{tot}}$ along the line $x_2 = 4L/5$ from DFP and ESMC simulations of the lid-driven cavity flow.}}
\label{fig:lid_driven_over_line}
\end{figure}

}

\subsection{\label{computational_cost} Computational Cost}
\begin{figure}
    \centering
        \centering
                \includegraphics{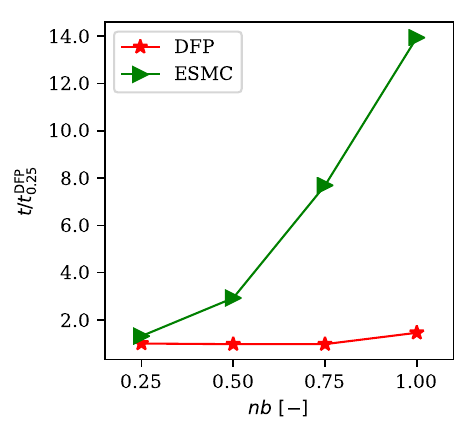}
    \caption{Execution time of ESMC and DFP normalized by the execution time of DFP at $nb=0.25$.}
    \label{fig:time.pdf}
\end{figure}

\noindent One of the main motivations behind the Fokker-Planck approximations of the Boltzmann or Enskog operators, is to reduce the computational complexity of the corresponding simulations. While a fair comparison between computational costs of different methods requires a separate study, here we provide how the costs of DFP and ESMC simulations scale with respect to the number density.
%One of the motivations for developing Fokker-Planck equation for dense flow (DFP) was computational advantage one should expect compared to DSMC based methods. Complexity of  DSMC based methods depends on the probability of collisions which is a quadratic function of density. The computational cost increases in case of ESMC furthermore since it requires a search for colliding pair from neighboring cells as well.
In order to study the effect of density on the computational cost, the equilibrium box problem mentioned in Section~\ref{sec:box} is deployed for $nb\in\{0.25, 0.5, 0.75, 1.0\}$.
%In this study, single core with one thread of  Intel(R) Xeon(R) CPU X7542  was utilized for all computations.
%\begin{table}[]
%\centering
%\caption{Performance and memory of Intel(R) Xeon(R) CPU X7542.}
%\label{tab:cpu}
%\begin{tabular}{c|c}
%Max Turbo Frequency  & 2.8 GHz  \\ \hline
%L2 cache & 256 K\\ \hline
%L3 cache &  18 MB  
%\end{tabular}
%\end{table}
%
As shown in Fig.~\ref{fig:time.pdf}, the ESMC cost sharply increases with respect to the number density. Yet, the complexity of the proposed DFP exhibits little or no correlation against the gas density. That is not surprising since the evolution of particles comprised in the DFP model are decoupled from each-other and thus the resulting cost only depends on the number of particles; not the number density.%This confirms that DFP can be considered for highly dense gases as an efficient alternative to ESMC at expense of losing accuracy.
\section{\label{sec:conclusion}Conclusion and Outlook}
\noindent The Fokker-Planck approximation of collision operators offers computationally desirable features for flow simulations, particularly at large densities and low Knudsen numbers. This comes from the fact that in the FP based models, the effects of intermolecular collisions are expressed through continuous stochastic processes which lead to an independent evolution of particles. Hence the computational cost remains constant against the density and the Knudsen number. \\ \ \\
\noindent In this study, the DFP model was devised as an approximation of the Enskog equation. The proposed cubic structure of the velocity and position evolutions leads to a moment system of the DFP model consistent with the one arising from the Enskog operator, up to the heat fluxes. The performance and accuracy of the DFP model was then evaluated for canonical test cases including Couette, Fourier {\color{myred}and lid-driven cavity} setups. Overall good agreements with respect to ESMC simulations could be observed. Future studies will focus on the extension of the DFP model for phase transitions and liquid flows, where the long range interactions will be included through the Vlasov-Enskog operator subject to the mean field limit.

\section*{acknowledgments}
\noindent The authors would like to acknowledge the useful comments from Profs. M. Torrilhon and P. Jenny.

\bibliographystyle{unsrt}%{siam}
\bibliography{My_Collection}
\end{document}